\documentclass[sigconf]{acmart}
\AtBeginDocument{%
  \providecommand\BibTeX{{%
    \normalfont B\kern-0.5em{\scshape i\kern-0.25em b}\kern-0.8em\TeX}}}

\setcopyright{acmlicensed}
\copyrightyear{2025}
\acmYear{2025}
\setcopyright{cc}
\setcctype{by}
\setcctype{by}
\acmConference[IUI '25]{30th International Conference on Intelligent User
Interfaces}{March 24--27, 2025}{Cagliari, Italy}
\acmBooktitle{30th International Conference on Intelligent User Interfaces (IUI
'25), March 24--27, 2025, Cagliari, Italy}\acmDOI{10.1145/3708359.3712086}
\acmISBN{979-8-4007-1306-4/25/03}

\usepackage{fontawesome5}
\usepackage{wrapfig}
\definecolor{pink_style}{RGB}{254,216,253}
\definecolor{table_color_1}{RGB}{220, 208, 255}
\definecolor{table_color_2}{RGB}{236, 235, 189}
\definecolor{table_color_3}{RGB}{250, 218, 221}
\definecolor{table_color_4}{RGB}{175, 238, 238}
\definecolor{table_color_5}{RGB}{255, 239, 213}

\usepackage{booktabs}
\usepackage{multirow}
\usepackage{graphicx}
\usepackage{subcaption}
\usepackage{hyperref}
\usepackage[framemethod=TikZ]{mdframed}
\usepackage{framed}
\usepackage{xspace}
\usepackage{enumitem}

\begin{document}

\title{NoTeeline: Supporting Real-Time, Personalized Notetaking with LLM-Enhanced Micronotes}

\author{Faria Huq}
\authornote{Both authors contributed equally to this research.}
\email{fhuq@cs.cmu.edu}
\affiliation{%
  \institution{Carnegie Mellon University}
  \city{Pittsburgh}
  \state{Pennsylvania}
  \country{USA}
}
\author{Abdus Samee}
\authornotemark[1]
\email{1805021@ugrad.cse.buet.ac.bd}
\affiliation{%
  \institution{Bangladesh University of Engr. \& Tech.}
  \city{Dhaka}
  \country{Bangladesh}
}
\author{David Chuan-en Lin}
\email{chuanenl@cs.cmu.edu}
\affiliation{%
  \institution{Carnegie Mellon University}
  \city{Pittsburgh}
  \state{Pennsylvania}
  \country{USA}
}
\author{Xiaodi Alice Tang}
\email{xiaodita@cs.cmu.edu}
\affiliation{%
  \institution{Carnegie Mellon University}
  \city{Pittsburgh}
  \state{Pennsylvania}
  \country{USA}
}
\author{Jeffrey P. Bigham}
\email{jbigham@cs.cmu.edu}
\affiliation{%
  \institution{Carnegie Mellon University}
  \city{Pittsburgh}
  \state{Pennsylvania}
  \country{USA}}

\newcommand{\todo}[1]{\textcolor{red}{#1}}
\newcommand{\revision}[1]{\textcolor{black}{#1}}
\begin{abstract}
Taking notes quickly while effectively capturing key information can be challenging, especially when watching videos that present simultaneous visual and auditory streams. Manually taken notes often miss crucial details due to the fast-paced nature of the content, while automatically generated notes fail to incorporate user preferences and discourage active engagement with the content. To address this, we propose an interactive system, \textit{NoTeeline}, for supporting \textit{real-time}, \textit{personalized} notetaking. Given `\textit{micronotes}', \textit{NoTeeline} automatically expands them into full-fledged notes using a Large Language Model (LLM). The generated notes build on the content of micronotes by adding relevant details while maintaining consistency with the user's writing style. In a within-subjects study (n=12), we found that \textit{NoTeeline} creates high-quality notes that capture the essence of \revision{participant} micronotes with 93.2\% factual correctness and accurately align with \revision{participant} writing style (8.33\% improvement). Using NoTeeline, participants could capture their desired notes with significantly reduced mental effort, writing 47.0\% less text and completing their notes in 43.9\% less time compared to a manual notetaking baseline. Our results suggest that NoTeeline enables users to integrate LLM assistance in a familiar notetaking workflow while ensuring consistency with their preferences\revision{—providing an example of how to address broader challenges in designing AI-assisted tools to augment human capabilities without compromising user autonomy and personalization.}
  
\end{abstract}



\newcommand{\tool}{NoTeeline\xspace}

\begin{teaserfigure}\includegraphics[width=\textwidth]{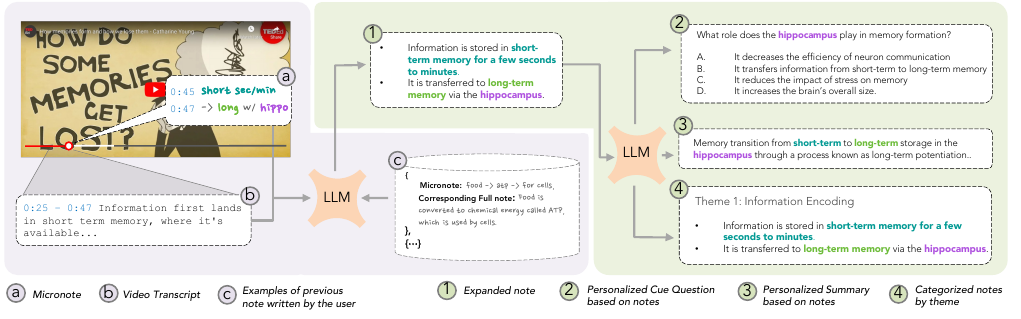}
  \centering  
  \caption{\tool facilitates on-the-fly notetaking while watching videos by expanding user-defined micronotes into richer-context notes written in their writing style.
  While watching a video, the user jots down quick keypoints (a). \tool retrieves the transcript at the timestamp (b). In addition, \tool references a small collection of the user's previous notes (c). Given (b) and (c), \tool automatically expands the micronote into a complete note in their writing style (1). Inspired by the Cornell notetaking system \cite{Saran2022AnIT}, \tool generates review questions (2), a brief summary (3), and groups notes into key themes (4) to assist users in note review.}
  \Description{}
  \label{fig:teaser}
\end{teaserfigure}


\maketitle

\section{Introduction}
Effective notetaking is crucial for capturing and organizing information. 
Whether in academic, professional, or personal settings, notes help individuals track important points, revisit information when needed, and maintain focus on the task at hand without getting overwhelmed by the flow of information. However, notetaking can be particularly challenging in a fast-paced environment like watching information-dense videos. 
Users may struggle with the high mental demands, forcing them to divide their attention between absorbing new content and documenting it \cite{Piolat2005CognitiveED, JANSEN2017223, workingmemperf}. Users frequently pause the video to write notes, disrupting the natural flow of content engagement \cite{VideoSticker} while forcing them to dedicate mental and physical effort to the mechanical act of notetaking. The inability to pause (e.g. in some live-streamed content or content on platforms that lack full media controls) adds even more pressure, often resulting in incomplete notes that may be difficult to interpret later \cite{mccoy2020gen, Kafipour2021ProcrastinationAI, flanigan2020impact}. 

\begin{quote}
    \textit{"The more content you try to capture during a lecture or a meeting, the less you're thinking about what's being said. You burn through most of your attention parroting the source." - Ryder Carroll \cite{carroll2018bullet}} 
\end{quote}

With the emergence of Large Language Models (LLMs,) there has been significant interest in developing automated notetaking tools from video transcripts (\textit{e.g.} NoteGPT \cite{notegpt_2024}, Otter \cite{otter}). \revision{While these tools may simplify notetaking, one common limitation is their lack of personalization.} Automatically generated notes are often generic and fail to address individual user needs: \revision{for instance, users may have specific writing style preferences \cite{biermann2022tool} or specific topics of interest}. As a result, these notes can feel impersonal and robotic, diminishing their usefulness for everyday use \cite{biermann2022tool}.



Alternatively, users may adopt shorthand notetaking strategies (\textit{i.e. micronotes}) to manage the high mental demand of fast-paced information consumption \cite{micronote}. They may jot down short keypoints containing shortened words or informational notations in order to write things down more quickly (see Figure \ref{fig:teaser}a). While micronotes help users keep up with the pace, they are often too vague to be useful during the later review. As a result, micronotes are often abandoned afterwards and not revisited \cite{Kleek2009NoteTS}. Thus, we observed a gap between quickly jotting down notes to capture information and having meaningful notes that can be revisited with clarity.


In this work, we leverage the practice of micro-notetaking and build an assistive note taking tool for videos. We introduce \tool\footnote{NoTeeline's name is inspired by Teeline\cite{wikipedia_contributors_2024}, a popular shorthand technique.}, an interactive notetaking assistance tool that aims to automatically expand user micronotes into complete notes in real-time, generated using their personal writing style.
We propose a novel pipeline that leverages video transcript for information extraction and a small handful of the user's previous note examples for preference modeling.
Through a within-subjects user study, we found that \tool helps users effectively capture notes with minimal disruption (e.g., reduced video pauses), capturing more content using fewer words and shorter time compared to a manual notetaking baseline. Participants rate the generated note expansions as high quality (average 6.0 on 7-point Likert scale, higher is better), align with the participants' micronote content (relatively high semantic similarity between micronotes and expanded notes measured with SBERT score \cite{sbert}), are consistent with participants' writing style (8.33\% improvement in chi-squared distance \cite{kilgarriff2001comparing}), and the notes contain low hallucination (93.2\% factual correctness score measured with HHEM score \cite{vectara2024}). \revision{For reproducibility, we provide our source code and user logs at: \url{https://github.com/oaishi/NoTeeline}}

\vspace{0.5pc}

\noindent In summary, this work makes the following contributions:
\vspace{-1pc}
\revision{
\begin{enumerate}
    \item We introduce \tool, a proof-of-concept notetaking tool built around the existing practice of \textit{micronotaking}. \tool leverages LLM pipeline to expand rough micronotes into full-fledged notes. \tool facilitates a fundamentally different approach to notetaking, centered on taking quick micronotes and letting LLM fill in the details, saving mental and physical effort.
    \item We conducted a within-subjects user study to evaluate the performance and utility of \tool. Our quantitative and qualitative results show that with \tool, users can take high-quality notes with higher efficiency, reduced disruption, and higher overall satisfaction, compared to a manual notetaking baseline. Users rated notes created using \tool with high consistency to their writing style (8.33\% improvement in chi-squared distance via SBERT score) and low hallucination (93.2\% HHEM score).
\end{enumerate}
}




\section{Related Works}


\subsection{Tools for Supporting Active Notetaking}
Active notetaking has been long practiced as an efficient strategy to store and retain information \cite{peper1978note, bauer2007selection}. Past research has introduced several interactive notetaking tools to support live-notes writing \cite{10.1145/191666.191697, Hinckley2012informal, Hinckley2007inksein, notestruct, tashman2011liquidtext, tsai2024gazenotercopilotedarnotetaking}, inking \cite{Forbus2011FORCSU, Sketchnote, VideoSticker}, information scaffolding \cite{bernstein2008evolution, Scraps} and note organization \cite{Spoerri2002SouvenirFN, meng2016hynote, subramonyam2020textsketch, He2014StructuringTU}. NoteLook supports multimedia note taking in meetings in an integrated conference room \cite{Chiu1999NoteLookTN}. VideoSticker proposes a visual notetaking tool for lecture videos utilizing object symbols to create multimodal, non-linear notes \cite{VideoSticker}. AQuA introduces a question-answering interface that helps users to watch software tutorial videos to get quick information \cite{yang2024aqua}. Livenote allows students to take lecture notes and annotate lecture slides cooperatively \cite{livenote}. NoteLink allows students to link their hand-written notes with the relevant video so that they can replay it easily \cite{Srinivasa2021NoteLinkAP}. Although these tools help users to organize and link source materials faster, users still have to write the whole note manually, often requiring additional laborious revisions and manual effort, causing a significant cognitive barrier \cite{Piolat2005CognitiveED}. On the contrary, \tool focuses on automatically generating the text of full notes from micronotes to shift effort from \textit{text notation} to capturing key ideas \textit{during active notetaking with users' own note points}. 

\subsection{Supporting Shorthand Writing}
Shorthand writing techniques have been popular with journalists for centuries \cite{mccay2021all}, to support higher writing speeds without loss of information. Inspired by their success, many digital inking software drew inspiration from shorthand techniques \cite{LEEDHAM2009}. This paved the way for a large body of research for \textit{gesture keyboard} \cite{word_gesture}. SHARK augments stylus keyboarding with shorthand gesturing by assigning a shorthand symbol to each word \cite{zhai2003shorthand, shark2}. Swipeboard encodes a sequence of gesture input into unique representations of each character to support faster typing in ultra-small devices such as smart watches \cite{swipeboard}. Both Swipeboard and SHARK report a higher writing speed compared to traditional keyboard. Padilla et al. introduce a deep learning-based methodology to automatically convert Gregg shorthand to English words \cite{9177452}. Although these tools show the potential of shorthand-based writing techniques in the modern era, they specifically focus on gesturing interaction. Oftentimes, they also require users to learn existing shorthand symbols. \tool provides users the flexibility to use their own abbreviated words and symbols \textit{without} requiring users to learn new gestures to use our system. To be best of our knowledge, \tool is the first tool to support micro-notetaking from videos and easy expansion of micronotes into full notes that are consistent with users' corresponding writing styles.


\subsection{Supporting Writing with LLMs}
LLMs have emerged as popular tools for writing assistance because of their versatile generation capabilities. A wide body of research has focused on enhancement of LLMs as writing assistant through building novel interaction tools \cite{gcompose, clark2018creative, freiknecht2020procedural, osone2021buncho, chung2022talebrush, ghai2021fluent} and improving generation capacity of these models \cite{mysore2023pearl, li2023automatic, context_helps, tang2017endtoendlearningshorttext}. Wordcraft allows users to write stories with LLM collaboratively with advanced features such as replacement or further prompting \cite{wordcraft}. Ramblers allows users to create voice memos and collaboratively use LLM to edit and expand on the voice memos to create a gist \cite{lin2024rambler}. Pearl proposes a retrieval-augmented pipeline that utilizes previous user-authored documents to generate texts of specific user style \cite{mysore2023pearl}. ExpaNet retrieves relevant documents to an input query and expands it using the retrieved documents \cite{expanet}. Outside academia, there is a growing number of commercially available tools for personal writing assistants as well \cite{producthunt2024}. While people frequently use these tools for assisting writing \cite{coauthor, wordcraft}, LLMs may struggle to preserve the style of experienced writers \cite{ippolito2022creative}. Current research is increasingly directed towards personalized writing aids \cite{biermann2022tool}. Compared to the above-mentioned tools, \tool is designed for robust user customization of note generation from start to end via control of references (i.e.: \textit{micronote}) and writing style. 

\section{User Study Setup}

Our study consists of two phases. First, we performed a needfinding study with 12 participants (\textsection \ref{sec:needfinding}, referred to as P1-P12) to understand the struggles they face in writing full notes in existing notetaking tools. 
Next, participants evaluated \tool in an hour session. (\textsection \ref{sec:eval}). The session was recorded in Zoom. 
We recruited our participants through mailing lists and announcement boards. The participants are situated in North America. They received a \$30 Amazon gift card for their participation. The study was approved by the IRB committee in our home institution. 

\section{Needfinding Study}
\label{sec:needfinding}
\subsection{Procedure} 
\revision{Few studies have explored how people use LLM-based note-taking systems alongside manually written notes. To better understand this area, including overall user perception and limitations, we conducted a needfinding study. Please refer to Appendix \ref{subsec:needfinding_ques} for the complete list of study questions.} 
Each study session was approximately 20-30 minutes long. \revision{Table \ref{tab:participant-table} shows the details for the participants who took part in the study. Our participants are in the age range of 20-30 ($\mu = 24.83, \sigma = 2.37$)}

The most common app used by participants was the built-in Macbook Notes app ($n = 9$). Many participants use pen-and-paper notes ($n = 7$) and later digitize them manually. In terms of AI-based tools, some participants used ChatGPT ($n = 4$) for summarizing meetings and video lectures. Others used Notion AI, Figjam AI, and Gemini plugin in Google Docs for specific use cases. 

\begin{table}[h]
\resizebox{\columnwidth}{!}{%
\begin{tabular}{cllcc}
\toprule
\textbf{ID} & \textbf{Nationality} & \multicolumn{1}{l}{\textbf{Age}} & \textbf{Gender} & \textbf{Occupation} \\ \midrule
P1 & South Korean & 30 & Male & Graduate Student (CS)\\
P2 & Chinese & 25 & Female & Undergraduate (HCI) \\
P3 & Chinese & 25 & Female & UX Designer \\
P4 & American & 22 & Female & \begin{tabular}[c]{@{}c@{}}Undergraduate \\ (Information Systems + HCI)\end{tabular} \\
P5 & Indian & 25 & Female & Graduate Student (HCI)\\
P6 & American & 21 & Female & \begin{tabular}[c]{@{}c@{}}Undergraduate\\ (Information Systems)\end{tabular} \\
P7 & South Korean & 25 & Female & \begin{tabular}[c]{@{}c@{}}Graduate student\\ (CS + Design)\end{tabular} \\
P8 & Indian & 25 & Female & \begin{tabular}[c]{@{}c@{}}Graduate Student\\ (Robotics + CS)\end{tabular} \\
P9 & Singaporean & 24 & Male & \begin{tabular}[c]{@{}c@{}}Undergraduate\\ (Fine Arts + HCI)\end{tabular} \\
P10 & American & 22 & Female & \begin{tabular}[c]{@{}c@{}}Undergraduate\\ (Civil Engr.)\end{tabular} \\
P11 & Indian & 27 & Male & Product Manager \\
P12 & American & 27 & Male & \begin{tabular}[c]{@{}c@{}}Software Engineer\end{tabular} \\ \bottomrule
\end{tabular}%
}
\caption{\revision{User study participant demographics.}}
\label{tab:participant-table}
\end{table}



\subsection{Insights and Observations} 
\label{subsec:insight}
Participants highlighted several challenges they face with their current notetaking process. We summarize these findings into three key themes through iterative rounds of coding guided by qualitative data analysis \cite{yin2015qualitative}:




\textit{C1. Users struggle to capture all desired information while taking notes from audio, video, and other real-time sources $(n = 7)$}.  Users mention resource constraints--time, attention, and mental load--in the practice of notetaking. P8 explains how their handwriting speed cannot match a professor's lecturing pace. P7 details the additional difficulties imposed by being a non-native English speaker. 

P11 finds it challenging to take notes while listening to podcasts, especially during physical activities like running on a treadmill. They describe the cumbersome process to jot down: \textit{`I have to slow down.. and then type it down in [a few words]... then I again need to speed up the treadmill and continue.'} The user emphasizes how the ad hoc nature of this process complicates effective notetaking, making it difficult to capture details in real time.



\textit{C2. Lack of agency in AI-assisted notetaking tools ($n = 6$)}. Users expressed the desire to maintain agency over AI-assisted notetaking tools. This is important given concerns about LLM `\textit{hallucinating}' non-existent content \cite{ji2023survey}. P1 says  \textit{`I would, at the end of the day, want some kind of source of truth... based on what the notes are going to be
'}. They reported to prefer using automated note features \textit{`as a compliment to whatever that [they're] using'} while \textit{`always going to take [their] notes no matter how'}, because of the \textbf{\textit{lack of trust}} in it. P2 further explains her thought -  \textit{`I hope it’s a tool for me... to help me get a better understanding of the video, not a tool for me to tell me what I should know because I still want myself to have the control of what the tool is doing.'} 


\textit{C3. Complex features and workflow of current tools. $(n = 7)$.} 
Participants highlighted that the complex features and \textbf{\textit{fragmented workflow}} of current notetaking tools often hinder usability. One common challenge involved managing two separate applications—one for watching videos and another for taking notes—which made the process cumbersome. They expressed a desire for having notetaking features within the video application, as this would facilitate a more seamless interaction, especially if the notes are linked to specific video segments to aid in later review.

Furthermore, participants reported that the overwhelming nature of feature-rich tools like Notion discouraged full utilization. As P5 described, \textit{`I don’t end up using the extent of features that Notion has to offer...  it’s so overwhelming .. so [I prefer] a simple interface but with the ability that as you go forward, you can start to explore more'.} In other words, the features are not well-situated or contextualized within the workflow, and presenting them all at once can overwhelm users, preventing them from fully utilizing the benefits.

\subsection{Design Goals}


Based on our needfinding study, iterative prototyping, and evaluation process, we propose three design goals for \tool.

\textbf{D1.} \textit{Micronotetaking as the way to take notes from videos to assist rapid writing.} The tool should support the quick creation of micronotes and be able to automatically expand it into a full note so that users do not need to write down all the details (C1). Even if the micronote includes grammatical errors or is extremely short, the system should be able to comprehend it. While transforming micronotes, it should be aware of the context, i.e., the topic being discussed in the video. The system should also support iterative editing of micronotes. 

\revision{\textbf{D2.} \textit{Allowing users to guide note generation using micronotes.} 
The system should enable users to guide note generation using micronotes, directing the LLM to focus on relevant video segments. LLM-expanded notes should exclude irrelevant or erroneous information (C2). Additionally, the system should adapt to each user's writing style, maintaining consistency.}


\textbf{D3.} \textit{Each component is linked to a specific stage of the notetaking process to enhance ease of use.} The system should link each component to a specific stage in the notetaking process—creation, editing, and revision. It should allow users to engage selectively with these features based on the current stage, providing ease of use throughout the notetaking process (C3) \cite{pea1987user, nielsen_2006}. 






\section{\tool}

\begin{figure*}
    \centering\includegraphics[width=0.9\linewidth]{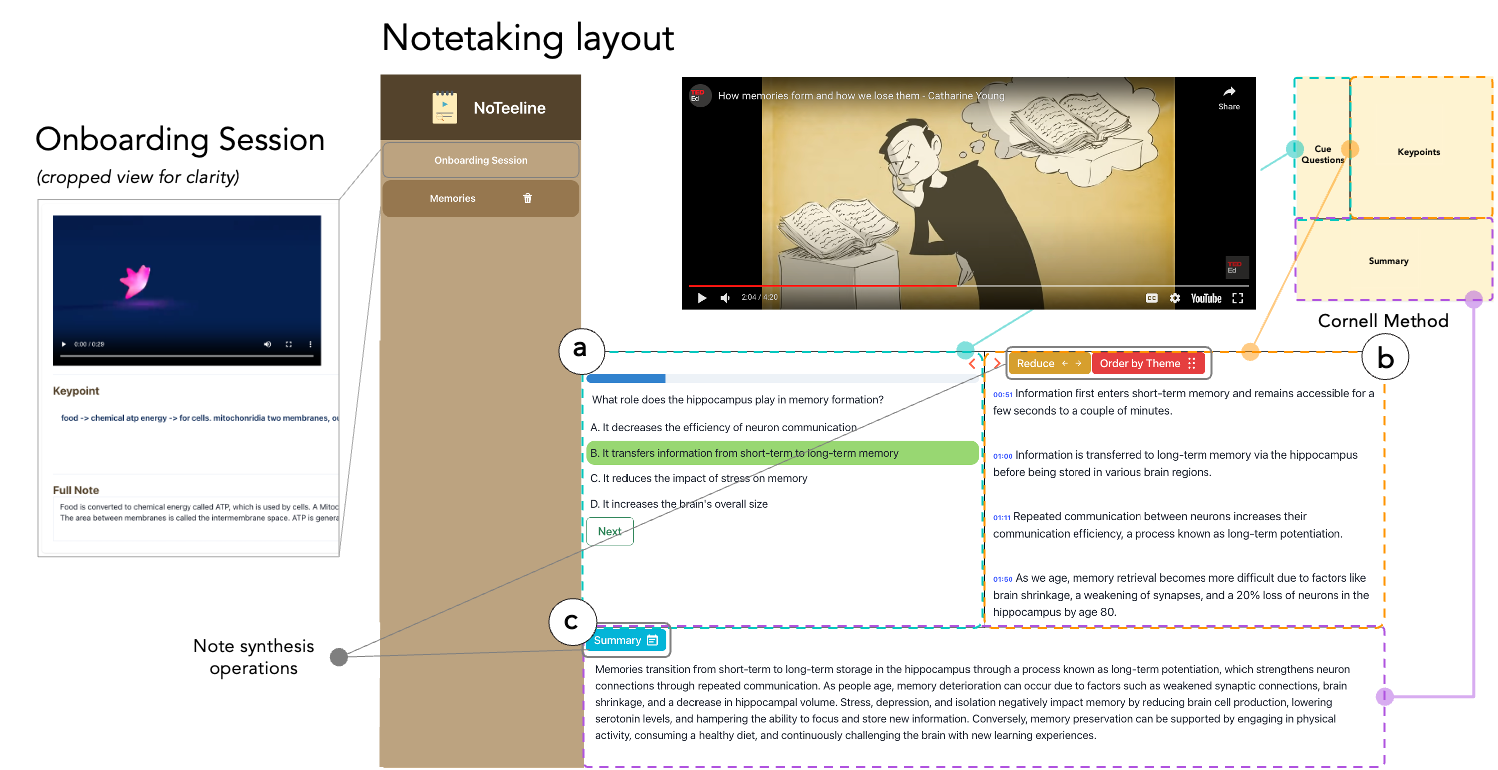}
    \caption{The \tool user interface consists of (a) the \textit{Cues} panel (outlined in blue) for review questions generated based on the notes, (b) the \textit{Notes} panel (outlined in orange) for user notetaking and note expansion, and (c) the \textit{Summary} panel (outlined in purple) for note summary. The layout is inspired by the Cornell method, as seen in the side-by-side comparison. The onboarding session, highlighted on the left, collects example notes from the user.}
    \Description{The \tool user interface consists of (a) the \textit{Cues} panel (outlined in blue) for review questions generated based on the notes, (b) the \textit{Notes} panel (outlined in orange) for user notetaking and note expansion, and (c) the \textit{Summary} panel (outlined in purple) for note summary. The layout is inspired by the Cornell method, as seen in the side-by-side comparison. The onboarding session, highlighted on the left, collects example notes from the user.}
    \label{fig:ui}
\end{figure*}

\subsection{Interface Components}

We will now describe individual components of \tool and how they facilitate our design goals.

\subsubsection{\textbf{Onboarding Session: Collecting Personalized Examples.}} We design an onboarding session to get reference examples of the user specific writing style (D2). Users watch three short clips (20-30 seconds each); and provide a keypoint as well as the corresponding full note for each clip (Figure \ref{fig:ui}). Once completed, the system will store these examples to use for note expansion afterward. 
Users need to complete this session \textit{only} once when they start using \tool for the first time. This session also gives users an initial experience of what a micronote would look like and what kind of full note to expect from the system. 

\subsubsection{\textbf{Notetaking Layout.}}
To better situate the interface around an effective workflow, we took inspiration from the Cornell method to construct the layout \cite{Saran2022AnIT}. We have three main regions -- 1) \textit{\textbf{Notes}}: the region for writing and organizing notes; 2) \textit{\textbf{Cues}}: auto-generated questions to review the note; and 3) \textit{\textbf{Summary}}: main takeaway of the note.

We also incorporated the video viewport on the top of the layout so that users can take notes while watching the video without needing to move between different desktop tabs. \tool progressively discloses the features as per users' current need for note writing, editing, or analyzing (D3). Figure \ref{fig:ui} illustrates the notetaking layout and its different components. 

\begin{figure*}[!ht]
    \centering\includegraphics[width=\linewidth]{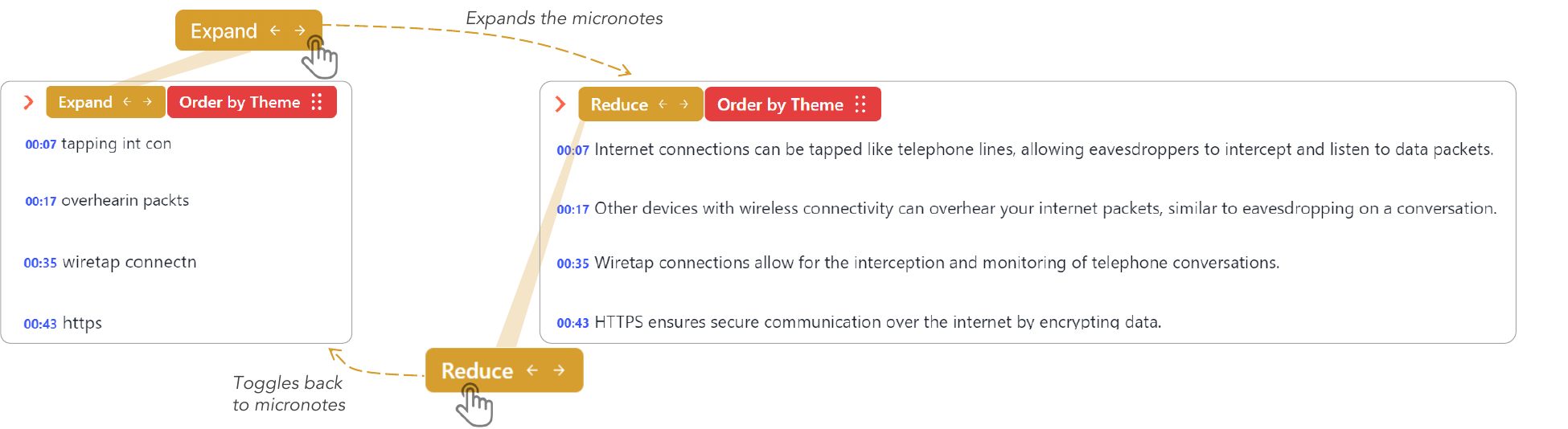}
    \caption{Micronote expansion. Users can toggle between the micronote and the expanded note by clicking the \textit{Expand}/\textit{Reduce} buttons.}
    \Description{Micronote expansion. Users can toggle between the micronote and the expanded note by clicking the \textit{Expand}/\textit{Reduce} buttons.}
    \label{fig:expansion_example}
\end{figure*}

\subsubsection{\textbf{Note Synthesis Operations.}} \label{subsec:synthesis}
We describe individual synthesis operations \tool supports and their mechanism. 

\textbf{\textit{Micronote Expansion.}} To construct the full note, we have an atomic \textit{expansion} operation. For \textit{expansion}, we utilize these pieces of information in the prompt: micronote, video transcript where the micronote was written, and reference examples from onboarding. 
The transcript helps to infer abbreviated symbols and fix potential typographical errors (D1) \cite{context_helps}. Reference examples help to understand users' writing style (D2). Figure \ref{fig:expansion_example} shows an example of micronote expansion.

\begin{figure*}
    \centering\includegraphics[width=\linewidth]{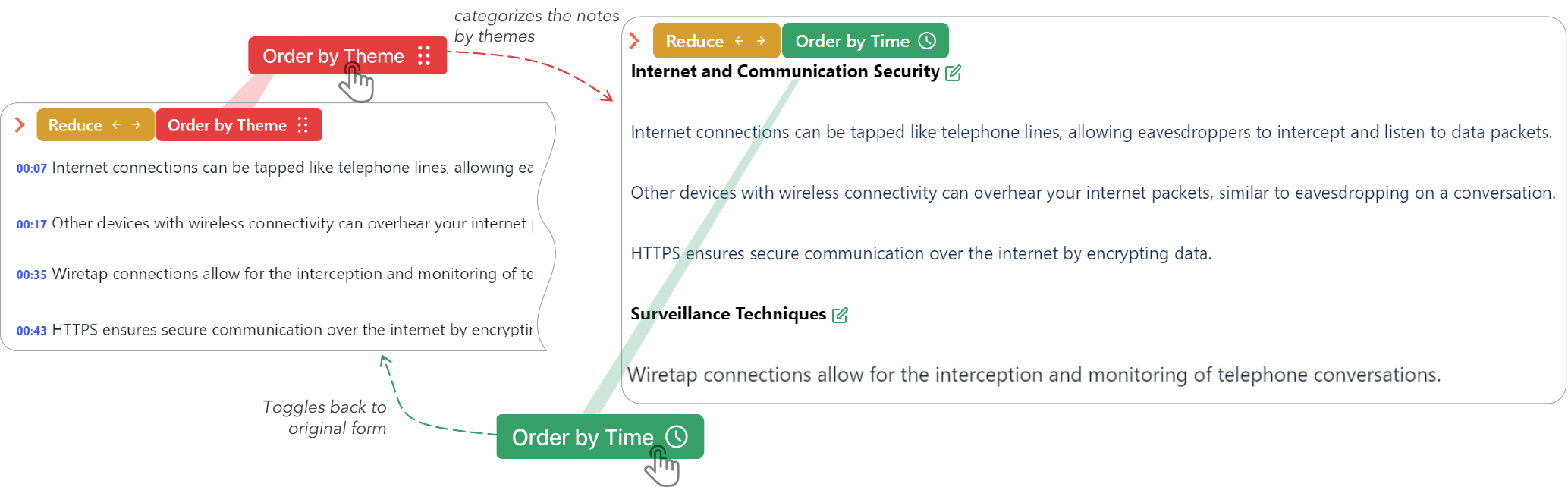}
    \caption{Note organization by themes (some texts have been cropped for clarity). Users can organize the notes by themes or toggle back to the original serial in which they took the notes.}
    \Description{Note organization by themes (some texts have been cropped for clarity). Users can organize the notes by themes or toggle back to the original serial in which they took the notes.}
    \label{fig:theme_example}
\end{figure*}

\textbf{\textit{Note Organization by Theme.}} To rearrange the note points by emerging themes, we have an atomic \textit{order by theme} operation. To \textit{order by theme}, we pass the generated notes to the LLM and instruct it to return the output in a structured, parseable format by using the few-shot prompting technique \cite{Graphologue, openai_2023}. We refer users to view the corresponding prompt Figure \ref{prompt-theme} in the Appendix to learn more about the structure we chose to use. Figure \ref{fig:theme_example} shows an example of how the note organization operation works.

\textbf{\textit{Note Modification.}} To edit individual micronotes or the expanded output from LLM and/or the automatically generated theme name (D3), users can click on the \includegraphics[height=0.8em]{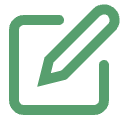} button next to each point.

\textbf{\textit{Note Rearrangement.}} To enable users rearrange their notes (D3), we have a \textit{drag and drop} operation. Users can drag individual note points and reposition them under a different theme using this operation. Users can also revert the note to its original serial by \textit{order by time} operation.

\textbf{\textit{Cue Question Generation.}} To enable users review their notes, we have a \textit{Cue Question Generation} operation, which generates 5 questions on-the-fly. If users want new questions, they have the option to initiate \textit{Regeneration} operation, which generates five new questions. The cue column is hidden by default to provide more space for writing to the users (D3). Cue question is generated in a similar structured fashion to \textit{order by theme} operation. Please view the corresponding prompt (Figure \ref{prompt-cue}) in the Appendix for details.


\textbf{\textit{Summary Generation.}} To generate a personalized summary, we have a \textit{Summarize} operation. The generated summary needs to reflect the user's note as well as the video context (D2). Hence, we passed the full video transcript and expanded the note to LLM for the summary. 


\subsection{System Walkthrough}

We present a hypothetical scenario to illustrate how users might interact with \tool. Kate is preparing a presentation on `\textit{Ocean Pollution}.' She uses \tool to quickly go through a long 40-minute and scaffold the information for her slide contents. She starts by completing the onboarding session and opens a new note named `\textit{Ocean Pollution Presentation}.' As she starts watching the video, Kate encounters a segment explaining the drastic increase in plastic pollution over the last decade. She quickly types "\textit{plastic pol. $\rightarrow$}" as a micronote. Once the video concludes, she presses the \includegraphics[height=0.8em]{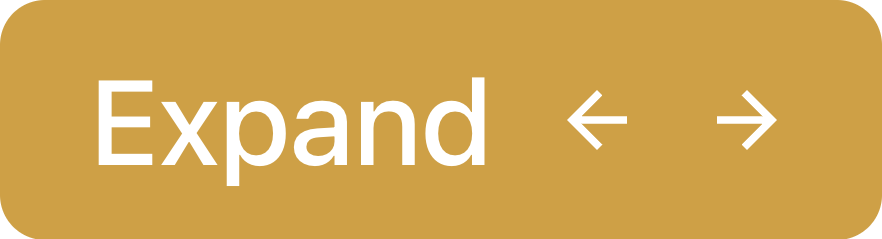} button, and \tool elaborates her note into a detailed point, including the specific statistic: "\textit{Plastic pollution in \textit{oceans} has increased by 200\% in the last 10 years.}" This expansion allows Kate to easily integrate precise details into her presentation later. 

Now, Kate needs to organize her information into themes. She clicks on \includegraphics[height=0.8em]{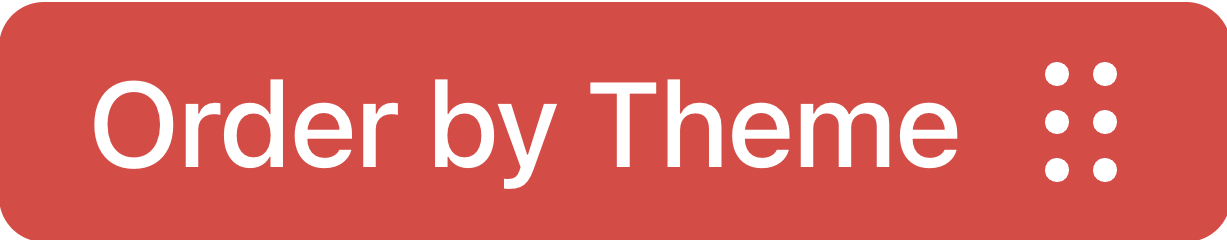} and it organizes her notes into three themes - `\textit{Causes of Ocean Pollution}', `\textit{Effects on Marine Life}', and `\textit{Conservation Efforts}'. While she is happy with the themes, Kate decides to further personalize the themes. She notices a point 'biodegradable alternatives to plastic' that would be better suited under `\textit{Conservation Efforts}'. Kate drags it under it and renames the theme to `\textit{Sustainable Solutions}' to better align with her presentation structure.

Feeling content with the organization of her notes, Kate clicks on \includegraphics[height=0.8em]{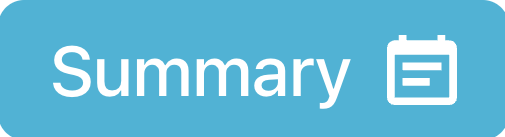} to generate a personalized \textit{TL; DR} summary for her conclusion. Finally, Kate wants to anticipate potential questions from her audience. She opens the \includegraphics[height=0.8em]{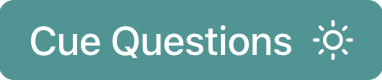} panel. 
Attempting to answer from memory, Kate correctly addresses the first question but stumbles on the second, prompting her to revisit her notes for a quick recap. 
Satisfied with her understanding and the comprehensive coverage of her topic, she confidently moves on to creating her presentation using her notes from \tool.

\subsection{\revision{Prototype Implementation}}
\label{sec:pipeline}

\revision{\tool is implemented as a web app using React.js and Chakra UI. A NodeJS server handles the API requests to the LLM. We used \texttt{youtube-transcript-api}, a Python package to extract the transcript from youtube videos.} LLM-based operations and text manipulation are performed using the OpenAI model, \texttt{GPT-4-Turbo} with hyperparameters set to temperature = 0.5, seed = 1. \revision{To reduce the randomness in API response, we used fixed model checkpoints with a predefined seed value.} We chose \texttt{GPT-4} over other LLMs due to its high-quality output and support for a longer context window, especially suitable for longer videos \cite{openai_2023}. \revision{On average, OpenAI takes $2.00$ seconds to expand all the micronotes in a single video (Please refer to table \ref{tab:time-pid} to view more details)}. 
We store all the data \revision{(\textit{e.g.} note points, transcription, summary ar onboarding examples)} in Chrome localstorage. 
\section{User Evaluation}
\label{sec:eval}
\label{subsec:study_proc}
With our user evaluation, we aimed to explore the following research questions:
\begin{enumerate}[label=\textbf{RQ\arabic*.}]
    \item Can \tool help users take notes efficiently with reduced disruption? 
    \item Can \tool generate note expansions that align with the user's micronote and personal writing style?
\end{enumerate}

\textit{\textbf{Participants.}} We invited 12 participants from our needfinding interviews to participate in a user study. Participants used both \tool and a baseline to capture notes and evaluate their performance.  They were divided into four subgroups to counterbalance and mitigate any ordering bias. Table \ref{tab:controlgroup} shows the distribution of participants across these subgroups. Participants were not exposed to information regarding the NoTeeline system before the user study.

\textit{\textbf{User Study Procedure.}} 
We first gave an overview of the study procedures to the participants. Participants were given the \tool web app link to browse from their laptops. They began with the onboarding session where we collected their personalized writing examples (\textsection\ref{appendix:onboarding_video}). After onboarding, participants were given two videos (\textsection \ref{subsec:vid_select}) to watch and take notes under the two study conditions. To prevent potential fatigue, participants were given a 5-minute break in between each video. After each video notetaking session, users were asked to complete a questionnaire \cite{tlx, sus, use}. Finally, we asked participants to try out \tool with a video of their choice. Each session was one hour long.


\textit{\textbf{Baseline.}} 
To compare \tool with a baseline, i.e., a basic notes app, we used a simplified view of our interface by hiding all advanced features (i.e., the synthesis functions described in \textsection \ref{subsec:synthesis}). This setting mimics the standalone Macbook Notes app, which most of the participants reported using. We also want to compare the utility between the personalized summary in \tool and the automatically generated summary in existing tools. Hence, we kept the summary feature in the baseline which is similar to NoteGPT \cite{notegpt_2024}, i.e., a summary that is \textit{not} conditioned on user notes, rather just summarizes the video transcript.

\section{Results and Findings}
\begin{figure}[!ht]
    \centering
    \hspace{-2pc}\includegraphics[width=\linewidth]{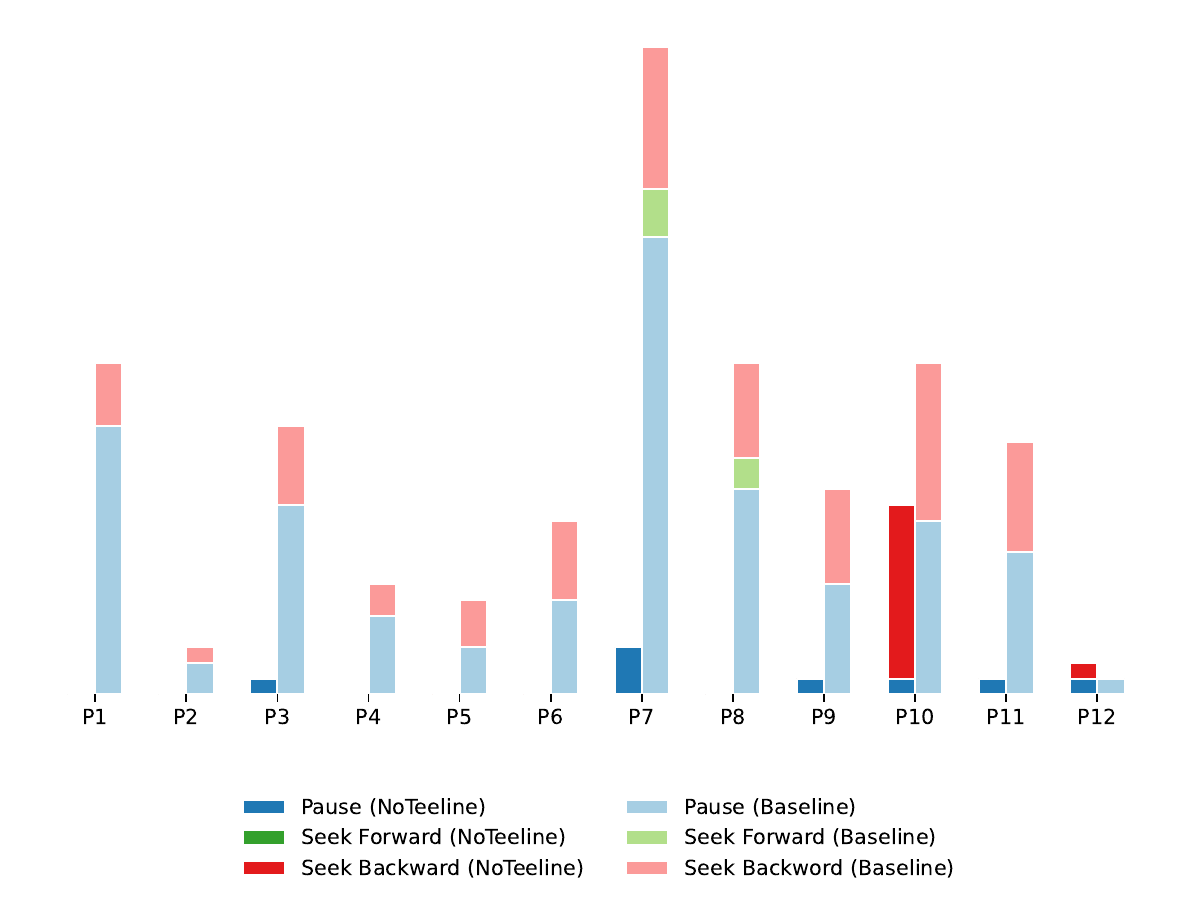}
    \caption{Distribution of pause and skips used by each participant in \tool (left) and Baseline (right). In \tool, the count of pause and skip is significantly less than the baseline. An empty bar depicts that the user did not pause or skip at all.} 
    \Description{Distribution of pause and skips used by each participant in \tool (left) and Baseline (right). In \tool, the count of pause and skip is significantly less than the baseline. An empty bar depicts that the user did not pause or skip at all.} 
    \label{fig:video_log}
\end{figure}

\label{sec:discussion}
We describe the results of our user evaluation, both quantitatively and qualitative showing the effectiveness of \tool in notetaking. 


\subsection{\tool Helps Users Write Notes Efficiently with Reduced Disruption (RQ1)}

\begin{figure*}
\centering
\begin{subfigure}[b]{0.33\linewidth}
  \includegraphics[width=\textwidth]{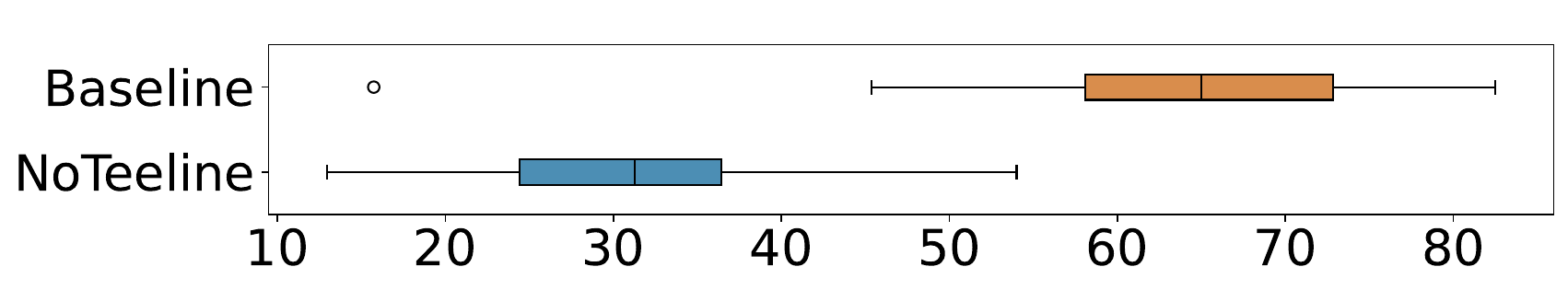}
  \caption{Average length of note point.}
  \label{fig:image1}
\end{subfigure}
\hfill 
\begin{subfigure}[b]{0.33\linewidth}
  \includegraphics[width=\textwidth]{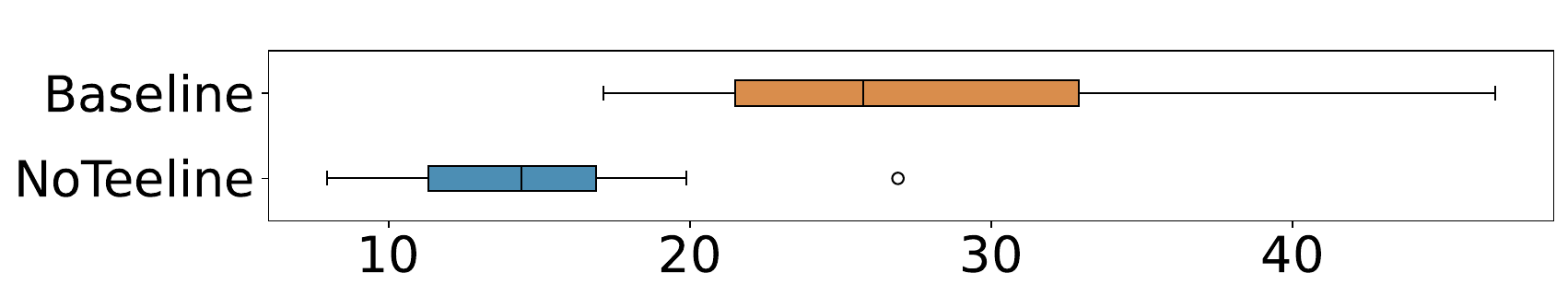}
  \caption{Average time per note point.}
  \label{fig:image3}
\end{subfigure}
\hfill
\begin{subfigure}[b]{0.33\linewidth}
  \includegraphics[width=\textwidth]{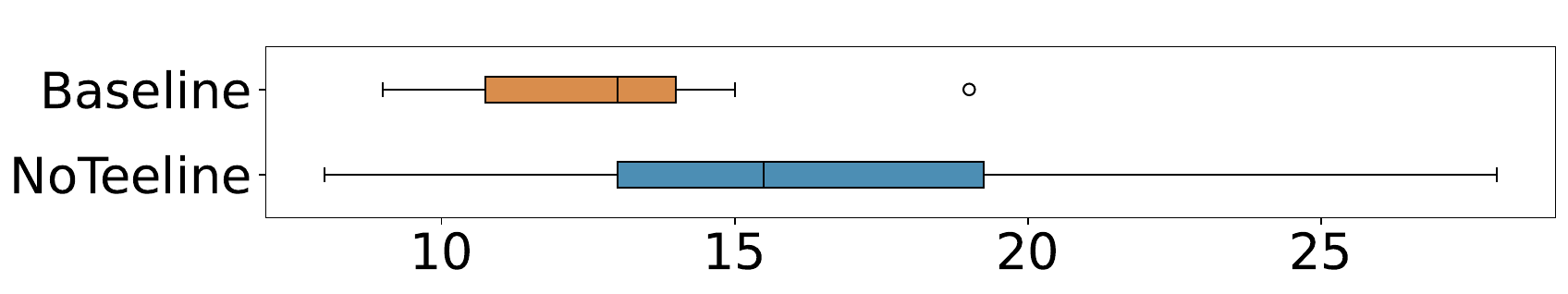}
  \caption{Total count of note point.}
  \label{fig:image2}
\end{subfigure}
\caption{Comparison of (a) note length, (b) writing time and (c) count for each participant between \tool and baseline. The majority of the users took the higher amount of notes in \tool in less time.} 
\Description{Comparison of (a) note length, (b) writing time and (c) count for each participant between \tool and baseline. The majority of the users took the higher amount of notes in \tool in less time.} 
\label{fig:threefigures}
\end{figure*}

\textit{Users efficiently captured the notes with less text in \tool.}
We report the average length of notes written by users in \tool ( $\mu = 32.11, \sigma = 12.33$), shorter than that in baseline ($\mu = 60.64, \sigma =16.94$) (Figure \ref{fig:image1}).

\textit{Users took a shorter time to write notes in \tool.}
Compared to baseline ($\mu = 27.79, \sigma = 8.25$), the notetaking time was 43.9\% shorter in \tool ($\mu = 15.59, \sigma = 5.4$) (Figure \ref{fig:image3}).

\textit{Users took more notes in \tool.} 
The total number of notes is more in \tool ($\mu = 18.75, \sigma =5.69$) compared to baseline ($\mu = 14.75, \sigma =2.59$) (Figure \ref{fig:image2}).  

\textit{Users faced fewer disruption in watching the video in \tool.} Users paused or rewound less with \tool ($ \mu =0.56, \sigma = 1.86$) compared to baseline ($\mu = 4.94, \sigma = 5.97$) when they were writing notes (Figure \ref{fig:video_log}).

The results indicate that \tool enables users to capture notes more efficiently by reducing the amount of text they need to write to express their thoughts. This also helps users to deeply concrete without disrupting their focus by pausing or rewinding. P4 directly speaks to this advantage: \textit{`I felt like I had to.. take every note [in Baseline].. you saw me like pause and rewind... Whereas in  [\tool\unskip], I felt like I could just watch it all the way through and not have to rewind.'} The higher frequency of notes also shows that \tool does not hinder the notetaking process in any way, rather it helps them to write it with \textit{less effort}. 


\begin{figure}
\centering
\begin{subfigure}[b]{0.4\linewidth}
  \includegraphics[width=0.8\textwidth]{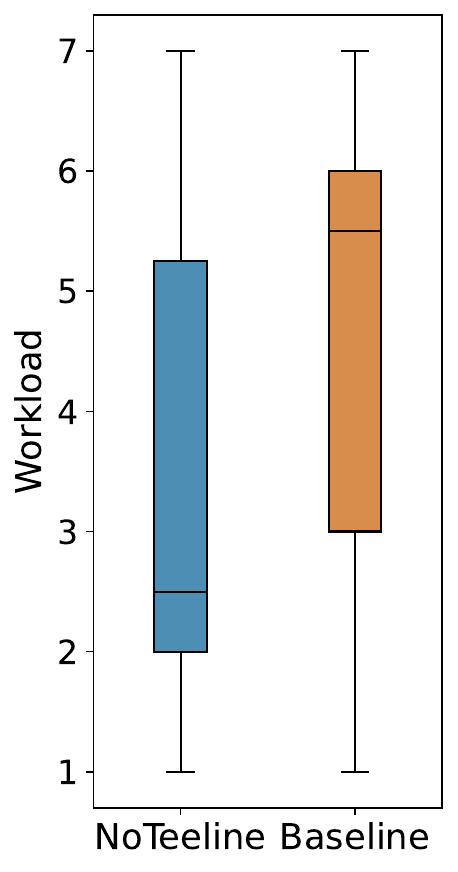}
  \caption{}
  \label{fig:tlx}
\end{subfigure}
\hfill
\begin{subfigure}[b]{0.4\linewidth}
  \includegraphics[width=0.8\textwidth]{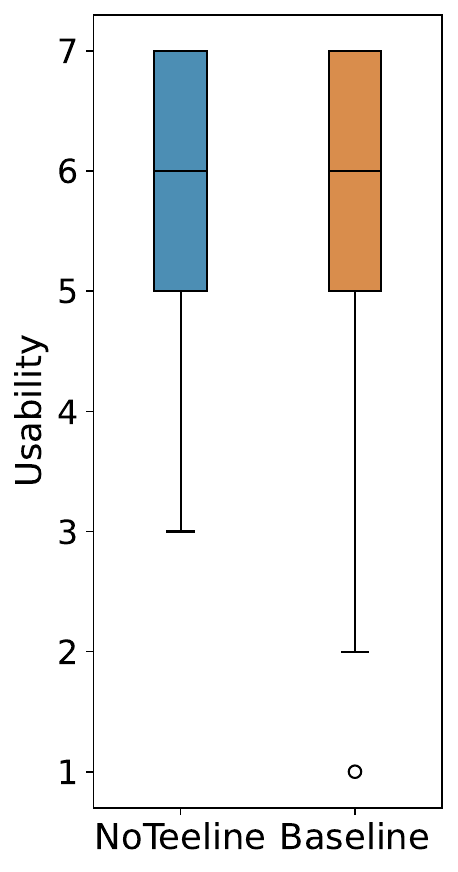}
  \caption{}
  \label{fig:sus}
\end{subfigure}
\caption{User study results (N=12). Boxplots from left to right: (a) workload measured with NASA TLX \cite{tlx} (7-point Likert
scale, lower is better), (b) usability measured with SUS \cite{sus} (7-point Likert scale, higher is better)} 
\Description{User study results (N=12). Boxplots from left to right: (a) workload measured with NASA TLX \cite{tlx} (7-point Likert
scale, lower is better), (b) usability measured with SUS \cite{sus} (7-point Likert scale, higher is better)} 
\label{fig:ratings}
\end{figure}

\subsubsection{Workload.} 
Using the Shapiro-Wilk test of normality, the differences between baseline and \tool for three of the four NASA-TLX questions \cite{tlx} on workload significantly deviate from a Normal distribution. Hence, we compare the differences in workload through a Wilcoxon One-tailed Signed-Rank Test \cite{neuhauser2011wilcoxon} with a Bonferroni-corrected \cite{weisstein2004bonferroni} significance level of 0.0125 per condition. Participants reported significantly lower results for Mental Demand, Effort, and Frustration ($p = 0.0123, p= 0.0024, p = 0.0047$). We did not find statistical significance for higher Performance (p = 0.0803). This may be attributed to users' general familiarity with notetaking as performance was already high in the baseline ($\mu = 5.58$ in the baseline versus a slightly higher average of $\mu = 6.25$ in \tool on 7-point Likert scale).

\subsubsection{Usability.} 
Usability was measured via an adapted version of SUS \cite{sus}) and yielded an overall average of 80.972, which corresponds to an A grade according to the Sauro–Lewis curved grading scale \cite{sauro2016quantifying}. We note that the average baseline SUS values are also high at 80.833. The high baseline usability may be influenced by the improved screen-space optimization in both conditions (C3). At a minimum,  our experimental features demonstrate high usability and do not seem to detract from the inherent usability of notetaking methods more familiar to users.

\subsection{\tool Generates High Quality Note Expansions (RQ2)}
\label{sec:nq}

\subsubsection{User evaluation.} During the user study, participants evaluated the quality of the automatically generated full note. \revision{Eleven out of 12 participants preferred the NoTeeline expanded note and summary compared to the automatically generated ones (Figure \ref{fig:likart_scale}).}


\subsubsection{Linguistic Property.} \revision{To measure linguistic quality and grammatical correctness,} we report the GRUEN~\cite{zhu2020gruen} of the expanded notes. The result (Table \ref{tab:gruen_scores} in Appendix) indicates that the points written using \tool ($ \mu = 0.84, \sigma = 0.07$) are of higher quality (i.e: less grammatical error, less redundant and more focused) than the baseline ( $\mu = 0.56, \sigma = 0.17$) ($p < 0.0001, t(11)=10.87$).

\subsubsection{Correctness} To ensure our automatically expanded notes are indeed correctly extracting the information from the transcript and not hallucinating any misinformation (C2), we compute the factual correctness of the generated note using HHEM \cite{vectara2024, zhang2023language, azamfirei2023large}. We noticed a Hallucination Rate = $6.8\%$ and a Factual Consistency Rate = $93.2\%$. This indicates that our generated notes are correct and consistent with the video content $93.2\%$ of the time.

\subsubsection{Consistency with Personal Writing style}
\begin{figure*}[h]
    \centering
    \begin{minipage}{0.5\textwidth}
        \centering
        \includegraphics[width=\linewidth]{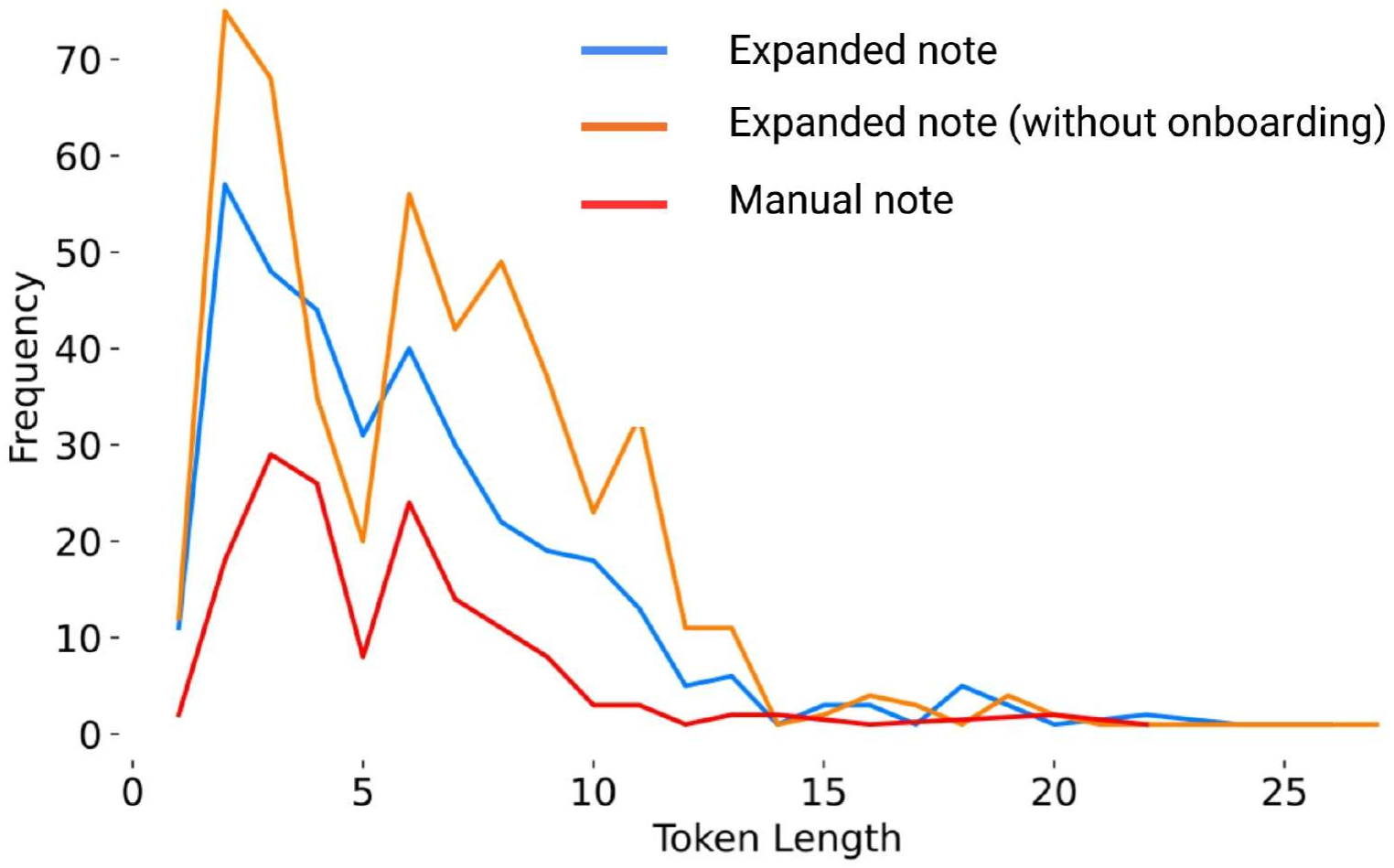}
        \caption{ Composition Characteristic curve \cite{mendenhall1887characteristic} for P9. The onboarding session examples help NoTeeline generate notes in a similar style that user would write themselves (the lower the gap between two curves, the higher the stylistic similarity)}
        \label{fig:mcurve}
    \end{minipage}
    \hfill
    \begin{minipage}{0.45\textwidth}
        \centering
        \begin{tabular}{llll}
        \toprule
        \multicolumn{1}{c}{\begin{tabular}[c]{@{}c@{}}\end{tabular}} & \begin{tabular}[c]{@{}l@{}}Expanded\\Note\end{tabular} & \begin{tabular}[c]{@{}l@{}}Expanded \\Note w/o \\ onboarding \end{tabular} & \multicolumn{1}{c}{\begin{tabular}[c]{@{}c@{}}Relative \\ Improvement\end{tabular}} \\ \hline
        P1 & 449.11 & 423.76 & {\color[HTML]{FE0000} -5.98\%} \\
        P2 & 335.21 & 375.47 & {\color[HTML]{009901} +10.72\%} \\
        P3 & 342.71 & 346.82 & {\color[HTML]{009901} +1.18\%} \\
        P4 & 297.96 & 336.86 & {\color[HTML]{009901} +11.55\%} \\
        P5 & 279.94 & 313.00 & {\color[HTML]{009901} +10.56\%} \\
        P6 & 423.84 & 428.98 & {\color[HTML]{009901} +1.20\%} \\
        P7 & 279.58 & 309.28 & {\color[HTML]{009901} +9.60\%} \\
        P8 & 417.74 & 402.64 & {\color[HTML]{FE0000} -3.75\%} \\
        P9 & 376.40 & 480.91 & {\color[HTML]{009901} +21.73\%} \\
        P10 & 287.53 & 353.18 & {\color[HTML]{009901} +18.59\%} \\
        P11 & 224.38 & 242.52 & {\color[HTML]{009901} +7.48\%} \\
        P12 & 496.38 & 579.93 & {\color[HTML]{009901} +14.41\%} \\ \hline
        avg. & 350.90 & 382.78 & {\color[HTML]{009901} +8.33\%}  \\ \bottomrule
        \end{tabular}%
        \caption{Chi-squared distance\cite{kilgarriff2001comparing} between the expanded notes from micronote and user's manual note. Examples of previous note collected during onboarding help us to achieve a better style similarity with the user's writing. {\color[HTML]{009901}Green} and {\color[HTML]{FE0000}Red} denote positive and negative relative performance change, respectively.}  
        \Description{Chi-squared distance\cite{kilgarriff2001comparing} between the expanded notes from micronote and user's manual note. Examples of previous note collected during onboarding help us to achieve a better style similarity with the user's writing. {\color[HTML]{009901}Green} and {\color[HTML]{FE0000}Red} denote positive and negative relative performance change, respectively.}  
        \label{tab:chi}
    \end{minipage}
\end{figure*}

To understand if the generated notes in \tool mirror the nuances of a user's personal writing style, we conduct a stylometric analysis \cite{stein2007intrinsic, li2023automatic, argamon2010structure} using Mendenhall’s composition characteristic curve \cite{mendenhall1887characteristic} and Kilgariff’s Chi-Squared \cite{kilgarriff2001comparing} methods.
These methods compare the choice of words and their distribution to infer the writing style. The closer the distribution curves are, the higher their style similarity. In Figure \ref{fig:mcurve}, we can see that the word distribution of expanded notes closely aligns with P9's hand-written note\footnote{For space limitation, we randomly show the characteristic curve of one participant.}, when we utilize the onboarding examples during note generation. The Chi-squared distance \cite{kilgarriff2001comparing} to the handwritten note is $350.90$ and $382.78$, respectively, with and without onboarding (lower is better) (Table \ref{tab:chi}). The results indicate that \tool achieves a better style match using the provided onboarding examples \cite{ning2024userllm}.

Participants also self-rated the consistency of expanded notes with their writing styles to be $\mu = 
5.5$ ( $ \sigma = 1.19$) on a scale of 1 to 7 (higher is better). P7 mentioned, `\textit{I have my own writing style... I use symbols like `$>$'... [NoTeeline] keeps using my style, so I loved it.}'



\subsubsection{Consistency of Micronote and its Corresponding Expanded Note}
\label{subsec:cons_mf}

To ensure our expanded notes are consistent with the main content of the micronotes, we measure the widely used sentence representation similarity, SBERT  \cite{sbert, ethayarajh2019contextual, gao2019representation}. We find that they have a 58\% similarity (avg. score = 0.58). We also examined the similarity of the expanded note and the transcript and found it to be 42\% (avg. score = 0.42). These results show that the expanded notes are more similar to micronotes than the transcript, indicating that our expanded notes prioritize the micronotes' contents and do not merely replicate the video transcripts, which will be undesirable.

\subsubsection{Consistency of Summary and Note}
\label{subsec:cons_sm}
To measure if the generated summary properly summarizes on the points mentioned in the note, we measure sentence representation similarity (SBERT) between the note and generated summary\cite{sbert, ethayarajh2019contextual, gao2019representation}. Compared to the baseline (i.e: NoteGPT \cite{notegpt_2024}), \tool summary is 11\% more relevant to the user note (Table \ref{tab:relevance}). Table \ref{tab:expand-summary} shows few examples of summary and expanded note pairs in the Appendix. 

\begin{table}[]
\resizebox{\linewidth}{!}{%
\begin{tabular}{llllllllllllll}
 & \textbf{P1} & \textbf{P2} & \textbf{P3} & \textbf{P4} & \textbf{P5} & \textbf{P6} & \textbf{P7} & \textbf{P8} & \textbf{P9} & \textbf{P10} & \textbf{P11} & \textbf{P12} & \textbf{avg.} \\ \toprule
\multicolumn{14}{c}{\textit{Consistency of Micronote, transcript \& corresponding expanded note (\textsection \ref{subsec:cons_mf})}} \\ \hline \\
\begin{tabular}[c]{@{}l@{}}Micronote \& \\ Expanded Note\end{tabular} & 0.6 & 0.41 & 0.73 & 0.75 & 0.63 & 0.58 & 0.55 & 0.6 & 0.48 & 0.43 & 0.57 & 0.63 & 0.58 \\ \\
\begin{tabular}[c]{@{}l@{}}Transcript \& \\ Expanded Note\end{tabular} & 0.47 & 0.41 & 0.42 & 0.42 & 0.42 & 0.39 & 0.42 & 0.42 & 0.44 & 0.46 & 0.39 & 0.41 & 0.42 \\ \\ \hline 
\multicolumn{14}{c}{\textit{Consistency of Summary \& Note in two conditions (\textsection \ref{subsec:cons_sm})}} \\ \hline \\
Baseline & 0.81 & \textbf{0.53} & 0.81 & 0.61 & 0.64 & 0.76 & 0.33 & 0.67 & 0.65 & \textbf{0.71} & 0.44 & 0.72 & 0.64 \\
NoTeeline & \textbf{0.87} & 0.51 & \textbf{0.9} & \textbf{0.63} & \textbf{0.91} & \textbf{0.77} & \textbf{0.67} & \textbf{0.8} & \textbf{0.68} & 0.7 & \textbf{0.76} & \textbf{0.82} & \textbf{0.75} 
\\ \\ \bottomrule
\end{tabular}%
}
\caption{Technical Evaluation Result (measured as \textit{sentence representation similarity} \cite{sbert}, higher is better)}
\vspace{-10pt}
\label{tab:relevance}
\end{table}

\subsubsection{Error Analysis}

\texttt{GPT-4} sometimes fails to generate the requested note. For example, P7 wrote a micronote to note down her thoughts that are not directly discussed in the video. The model replied with - \textit{`Please provide the transcript related to the keypoint so I can assist you in creating a note.'}. The occurrence of such error is 0.72\% (4 out of 559 total calls). 

\begin{table*}[ht]
\resizebox{\textwidth}{!}{%
\begin{tabular}{@{}ll@{}}
\toprule
\textbf{Micronote} & \textbf{Expanded note} \\ \midrule
RNNs are unrolled \colorbox{table_color_1}{l to r or opp} & Recurrent neural networks (RNNs) can be unrolled either from \colorbox{table_color_1}{left to right or from right to left}, encoding \colorbox{table_color_4}{linear locality}. \vspace{8pt} \\
find the average / adding vectors together & Average of vectors \colorbox{table_color_2}{-\textgreater} add vectors together \colorbox{table_color_2}{-\textgreater} find mean value. \vspace{8pt}\\
red \colorbox{table_color_5}{-\textgreater} all similar reds & Red color categorization \colorbox{table_color_5}{includes} various shades that are perceived as similar. \vspace{8pt}\\
Cites difficulties for \colorbox{table_color_1}{BVI} people & \begin{tabular}[c]{@{}l@{}}The speaker cites the difficulties faced by \colorbox{table_color_1}{blind and visually impaired} individuals, such as the \colorbox{table_color_4}{inability to independently} \\ \colorbox{table_color_4}{read books, magazines, or use computers and smartphones} without assistance.\end{tabular} \vspace{8pt}\\
sauce: \colorbox{table_color_3}{gehra?} & \begin{tabular}[c]{@{}l@{}}The sauce, known as \colorbox{table_color_3}{garum}, was a \colorbox{table_color_4}{staple in Ancient Greek and Roman cuisine}, valued for its strong umami flavor and \\ produced in large quantities throughout the Mediterranean.\end{tabular} \vspace{8pt}\\
\colorbox{table_color_1}{both} rich and umami & \colorbox{table_color_1}{Tomatoes, like fish sauce}, are rich in \colorbox{table_color_4}{glutamate} which gives them a strong umami flavor. \vspace{8pt}\\
css property \colorbox{table_color_5}{--\textgreater} native styling & Custom properties in CSS can be used to \colorbox{table_color_5}{create} native styling options for web design. \vspace{8pt}\\
\colorbox{table_color_1}{met} & The \colorbox{table_color_1}{Metropolitan Museum of Art} is located on Fifth Avenue and spans five blocks. \vspace{8pt}\\
))) \colorbox{table_color_5}{-} russian & In Russian, multiple closing brackets ")))" \colorbox{table_color_5}{indicate} \colorbox{table_color_4}{varying degrees of happiness} in a smiley face. \vspace{8pt}\\
pipe \colorbox{table_color_5}{--\textgreater} high to low pressure & Water in pipes \colorbox{table_color_5}{flows from} areas of high pressure to low pressure, similar to how electrons move from high to low voltage. \vspace{8pt}\\
equipment durability & \colorbox{table_color_4}{Cycling pumps} on and off to meet variable demand increases wear and tear, affecting equipment durability. \vspace{8pt}\\
slow - reducing fat mass. fast \colorbox{table_color_5}{-} reducing \colorbox{table_color_3}{muscule} mass & Slow dieting reduces fat mass, while fast dieting \colorbox{table_color_5}{leads to} \colorbox{table_color_3}{muscle} mass loss. \vspace{8pt}\\
\colorbox{table_color_1}{1} \colorbox{table_color_3}{enery} deficit & \colorbox{table_color_1}{Sam} creates an \colorbox{table_color_3}{energy} deficit by \colorbox{table_color_4}{consuming fewer calories and increasing exercise}, leading his body to break down glucose stores. \vspace{8pt}\\
loose fast\colorbox{table_color_5}{? \textless} find healthier life style & Losing weight quickly isn't ideal; \colorbox{table_color_5}{instead}, focus on discovering a lifestyle that promotes personal health and happiness. \\ \midrule
\multicolumn{2}{c}{\textbf{Applied Changes:} \colorbox{table_color_1}{\vphantom{X}\hspace{0.8em}}  Abbreviation Expansion \vphantom{X}\hspace{1.5em} \colorbox{table_color_4}{\vphantom{X}\hspace{0.8em}} Detailed Information Extraction \vphantom{X}\hspace{1.5em} \colorbox{table_color_2}{\vphantom{X}\hspace{0.8em}} Personal Writing Style Imitation \vphantom{X}\hspace{1.5em}  \colorbox{table_color_5}{\vphantom{X}\hspace{0.8em}} Personalized Symbolic Interpretation} \vphantom{X}\hspace{1.5em} \colorbox{table_color_3}{\vphantom{X}\hspace{0.8em}} Typo Fix\\ \bottomrule
\end{tabular}%
}
\caption{Example micronotes written by our participants and the corresponding expanded notes in \tool. The relevant portions in the micronote and expanded note are highlighted for readability. The final notes can properly expand the abbreviation as well as symbols, extract detailed information, mimic personal writing style, and fix typographical errors.}
\label{tab:example_notes}
\vspace{-10pt}
\end{table*}

\subsection{How Users Use \tool and Individual Features}
\subsubsection{\tool Supports Diverse Notetaking Patterns}

Table \ref{tab:example_notes} presents example micronotes our participants generated and their corresponding expanded notes generated by \tool. \newline \textit{\textbf{Patterns:}} Participants utilized a variety of patterns when constructing their micronotes. Symbol usage was frequent, with `->' commonly used to indicate a process. `?' was often employed to denote uncertainty or to highlight areas where participants sought further clarification. Abbreviations were widely used as well. For instance, P1 used the acronym '\textit{BVI}' to represent the '\textit{blind or visually impaired community}.' P2 employed the numbers '\textit{1}' and '\textit{2}' to refer to characters '\textit{Sam}' and '\textit{Felix}' from a video. Some participants shortened words by using the first letter or a few subsequent letters. For example, P8 wrote '\textit{l to r}' to represent '\textit{left to right}.' \newline\textbf{\textit{Operations.}} We observed that \tool supports five types of operations during note expansions. 1) \textit{Abbreviation Expansion}: Converts abbreviations into their full form; 2) \textit{Detailed Information Extraction}: extract missing information from transcript required for note expansion. 3) \textit{Personal Writing Style Imitation}: mimic the writing pattern users usually use in their writeup; 4) \textit{Symbolic Interpretation}: accurately decodes the symbols in user-specific writing style and integrates them into the expanded notes; 5) \textit{Typographical Error fix}: corrects spelling mistakes or missing letters in the text.



\subsubsection{\tool Supports a Variety of Content Domains}
During the freeform test (\textsection \ref{subsec:study_proc}), Participants reported and used \tool in a wide range of video categories. All 12 participants wished to continue using \tool in the future. P7 reported that they continued using \tool for annotating research interviews even after our study session. P9 mentions the desire for integration with a current Learning Management System, Canvas, used at their university. Most participants thought \tool would be most useful for long-form videos where there is a lot of information to capture. 
The top categories they reported are: 1. \textit{Educational Video Content ($n = 8$)}; 2. \textit{Professional Podcast and meetings ($n = 3$)}; 3. \textit{Thematic analysis of research videos ($n = 3$)}; 4. \textit{Personal Learning and Self-Education from how-to tutorials ($n = 2$)}; 5. \textit{Financial and Investment Learning ($n = 1$)}. 


Even when features were not immediately relevant to their current needs, participants reported potential uses for others or themselves at other time points in their lives. P12 says the cue questions would be useful for \textit{`my (their) friends at med school'}  where `\textit{the details get really granular and... the names of things can can make or break like the answer to a question.'} P6 mentions how, even though they are no longer a student, they would have `\textit{definitely want[ed] to use [\tool\unskip] with [their] lectures... freshman and sophomore year when classes were online.'}





\subsubsection{Comments on \tool's Individual Features}
\label{subsec:features}

\begin{figure}[ht]
    \centering
    \includegraphics[width=\linewidth]{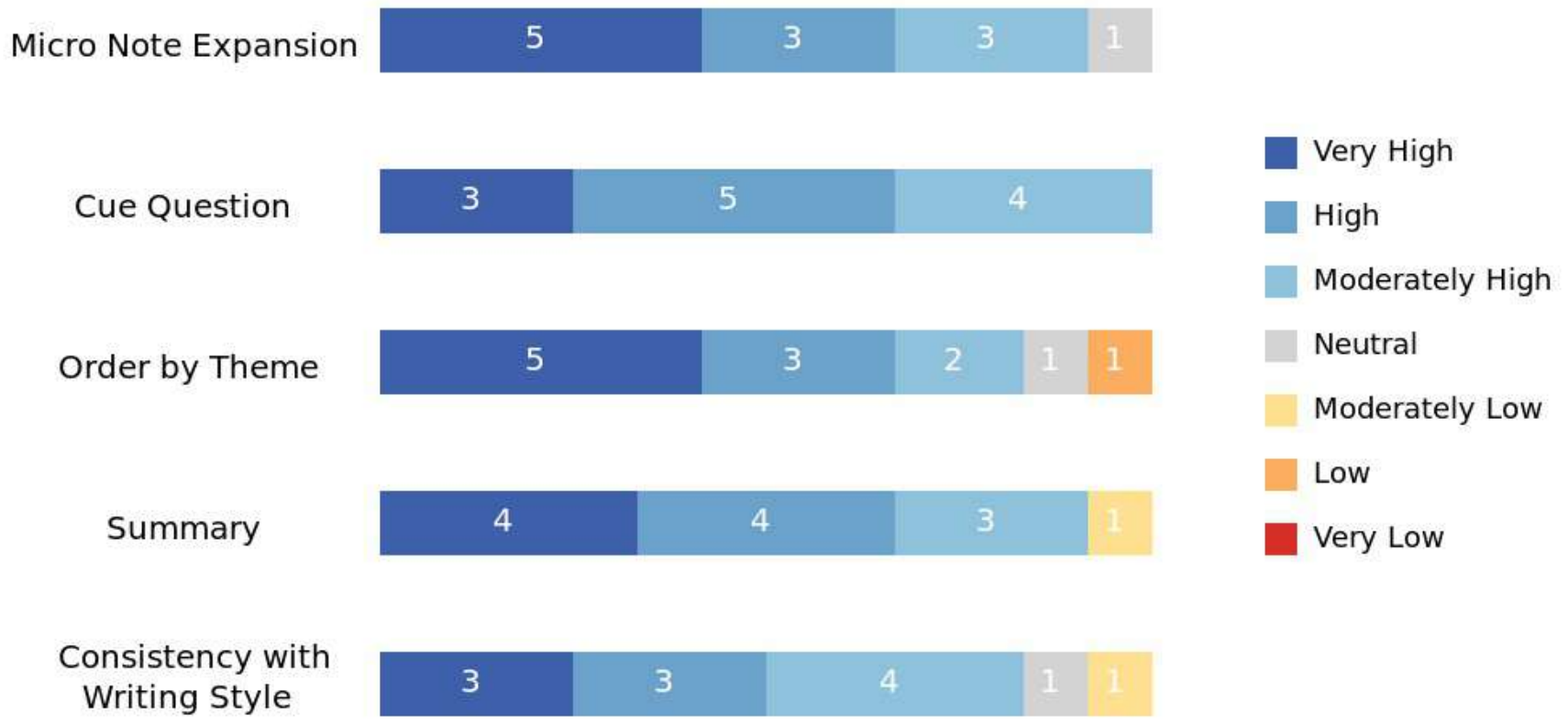}
    \caption{Participants’ responses to utility and usability of \tool interface and various features (7-point Likert scale, higher is better).}
    \Description{Participants’ responses to utility and usability of \tool interface and various features (7-point Likert scale, higher is better).}
    \label{fig:likart_scale}
\end{figure}

In figure \ref{fig:likart_scale}, we show participants rating for individual features in \tool ($\mu = 5.8, \sigma = 1.15$) (7-point Likert scale, higher is better). Overall, \tool was received positively by the users. 



\textbf{Impact of Note Expansion.} All 12 participants reviewed some aspect of the expanded note favorably. All critical feedback, present in four of the responses, focused on the potential for verbosity and redundancy in the expanded notes (refer to \textsection \ref{sec:nq} for demonstration of how \tool still improves upon the baseline for measures of redundancy). Still, nine reviews expressed appreciation for the quality/level of detail in the expanded note, and five reviews brought up that the expanded note captured details/context they would have otherwise missed, due to such reasons as cognitive and time constraints (e.g. P9: \textit{`\textbf{focusing on the keywords is less mentally draining and would be helpful in keeping my attention span for a longer period of time}}'; P6: `\textit{there’s like a lot of things that I heard I probably wouldn’t have to have time to type it all out, but it was able to expand on everything}'). Eight participants also specifically remarked that the expanded note reflected their writing style. P11 says  \textit{`If my note was seven words or eight words then it went like one and a half lines if mine was small and then it was also like not trying to bring in too much content so that I felt it was trying to match with my style of what I’m trying to put'.}


\textbf{Impact of Generated Themes.} 11 out of 12 participants reviewed some aspects of the generated themes positively. 
P8 believes the impact of Order by Theme would become more evident in longer videos. 
P10 discusses how it would have a greater impact in a video with more discursive content. In that regard, P11 expresses excitement about using the Order by Theme in such a context \textit{`where the conversation may not usually flow through on one topic'.} Five reviews specifically point out the utility of the Order by Theme feature for aiding the review, retrieval, and organization of content.


\textbf{Impact of Cue Questions.} 11 out of 12 participants reviewed some aspects of the cue questions positively, with critical feedback from five participants tending to focus on increased cue question complexity. P9 identified that multiple choice questions may not be applicable for \textit{`concepts [that] are more open to interpretation'}. Six participants specifically brought up positive impressions of the cue questions aiding in the video/note review process, four participants mentioned that they viewed the cue questions as a way to identify potential shortcomings in their notes, and three participants highlighted the perceived relevance of cue questions to the video content.


\textbf{Impact of Summary.} 11 out of 12 participants reviewed some aspect of the summary feature positively. Four responses brought up satisfaction with the summary's ability to capture the gist/important contexts from their notes. Three responses mentioned a preference for notes versus the generated summary for reviewing purposes: P1 prefers reviewing \textit{`small details'} while P3 and P6 find the length too long. However, opinion on length is not uniform, as five responses (P5, P7, P8, P10, P12) reflected positively on the length of the summary. Four responses described a desire for more formatting features, such as bolded or highlighted text.

\textbf{General Reception of AI Assistance.}
We conclude the thematic analysis with a general review of reception to NoTeeline's AI assistance during notetaking. Across the user study, ten participants specifically addressed the relationship between their labor and the AI features in NoTeeline impacting their workflows. P1 speaks to the `\textit{division}' of labor in which they capture the `\textit{smaller details}' while the `\textit{machine... take[s] over the higher level summary}.' P1, P3-P4, P6-P8, P10-P12 speak to how the automated features distributed throughout NoTeeline provide \textit{\textbf{enhancement or safeguard}} beyond their individual labor--whether capturing missed details/context, correcting typographic and grammatical errors or otherwise refining their writing, or surfacing unrecognized themes or shortcomings in their understanding of the material.

\section{Discussion}
\subsection{Summary of Findings}
Building upon the established concept of micronotes, \tool supports real-time notetaking from videos in a fast-paced environment. We found that \tool reduces the manual effort required for writing full notes, allowing users to focus on capturing core ideas rather than the tedious efforts of transcribing content verbose. Our study shows that users were able to capture a large amount of content in their notes with significantly fewer words and less time compared to manual notes. Our study also reveals that users are able to use different note taking patterns while \tool correctly decodes their micronote. All participants in the study preferred \tool over the baseline and envisioned themselves using it in the future. By providing users the ability to write their own notes while enabling LLM assisted automated in a personalized way, \tool achieves user trust and gives them a sense of agency.

\subsection{Supporting Uninterrupted Active Notetaking Experience}

While one might argue that spending more time to write individual notes is a better way to get a deeper engagement with the materials, and \tool might seem to deviate users from writing effective notes, we argue that \tool does not detract, but rather enhances the engagement in more meaningful ways. 

First, users are still responsible for initiating which parts of the material to take notes for and what information they want to capture, with \tool simply helping to transform their input into fuller structured notes more suitable for later review. In contrast to existing fully automated tools, users are actively involved in notetaking. 

Second, \tool reduces the disruption in watching videos caused when taking full notes (e.g., constantly having to pause the video) and freeing them from the mechanical task of writing. Thus, It helps maintain the uninterrupted flow of information. 

Third, \tool also allows deeper involvement in note analysis and revision stages. After the users have finished taking the notes, they can always come back to it and edit, reorganize, and revise based on their needs. Our system provides features to review notes by the cue question, to manually edit the automatically expanded notes, to jump to the location where the note was taken, and to organize the notes in their desired structure and themes. These features were inspired by the Cornell notetaking system with their long history of improving long-term knowledge retention \cite{Kornell2014-wt, Evans2019UsingTC, Yuniarti2018DevelopingSN}. All these factors altogether enable users to deeply engage with content in \tool.



\subsection{Insights for Human-AI Collaborative Notetaking Tools}




Based on our findings from the design and evaluation of \tool, we highlight three insights to inform the future design of human-AI collaborative notetaking systems. First, our findings reveal that the greatest benefit
of \tool is the affordance of the micro note and
its expansion. Being able to continue writing notes within their usual workflow while enjoying LLM assistance sparks interest. The ability to toggle back and forth between the original and modified notes enhances their overall experience by providing control over the content, a critical factor in fostering user trust \cite{jeung2023correct}. Future tools should support iterative processing, including automatic cleaning of grammatical errors, and offer both LLM-assisted and manual options for editing and organizing notes.

Second, collaborative notetaking systems should provide granular user control. In our needfinding study, users mention concerns about LLM hallucination and lack of control in AI-assisted tools (\textsection \ref{subsec:insight}). Prior Research shows that trust in human-machine teaming can be positively impacted through transparency \cite{alhaji2020engineering}. By giving users the ability to control precise moments to take notes and writing styles to maintain -- instead of relegating the generation task entirely to a software -- \tool can increase transparency in the notetaking process. 

Finally, an AI-assisted system should leverage \textit{implicit} guidance \cite{jannach2018recommending} from users to regulate its generation quality. In \tool, users would take micronotes as usual while \textit{implicitly} guiding the system about what to focus on in the generated note. Such support for implicit guidance does not require additional effort from the user such as writing custom prompts for their individual requests every time.




\section{Limitations and Future Work}



While \tool was generally positively received, there are some avenues for future work that we would like to explore. First, \tool enables users to take notes as they progress through the content, following a linear notetaking workflow \cite{meng2016hynote}. It does not directly provide support for writing non-linear notetaking (though does provide support for non-linear organization of the notes by theme). Second, an interesting future extension could be to incorporate \tool with a more visual notetaking tool such as  VideoSticker \cite{VideoSticker}, so that users can capture images and figures from videos within their notes.

While our study demonstrates that NoTeeline significantly reduces cognitive load and improves notetaking efficiency by minimizing writing effort and time, we did not measure the long-term effect of \tool on information retention. The primary goal of this work was to evaluate the system’s ability to facilitate efficient note-taking, not to assess how the generated notes impact user learning outcomes. This follows standard practice in prior research, as seen in works like Rambler\cite{lin2024rambler} and VideoSticker\cite{VideoSticker}.
Nonetheless, for future work, we could investigate how NoTeeline can influence information retention through a longitudinal study.


\section{Conclusion}
Active notetaking from videos remains challenging. While existing tools can automatically generate notes, they do not necessarily capture the details and stylistic nuances users care about. Our needfinding study reveals that users often do not prefer automatically generated notes due to concerns about accuracy and lack of customization. \tool leverages rich contextual information to generate more accurate and more personalized notes. Our user study reveals that \tool helps write notes faster with $93.2\%$ correctness and achieves closer stylistic characteristics to users' handwritten notes. We demonstrate the utility of \tool in a variety of usage scenarios users reported and provide design implications for future works in AI-assisted notetaking systems.

\section*{Societal Impact}
By helping users take notes faster, \tool has the potential to make notetaking more accessible for individuals who might otherwise struggle with factors such as language fluency (e.g., P7’s comments about difficulties of being a non-native English speaker) and physical disabilities. While there exist arguments about whether over-reliance on AI-generated content may lead to passive engagement or diminish personal organizational skills, we show how our tool exists to \textit{amplify, not override}, user engagement in the generation process. We also expect this to decrease the risk of AI biases that might otherwise occur in fully autogenerated notes without any sort of user engagement. 
\begin{acks}
\revision{
    We thank Dan Saffer and Chris Maury for their valuable feedback on the design of \tool, which contributed significantly to its development. We also appreciate the insightful comments and suggestions provided by Pranav Khapde, Nazmus Saquib, Riku Arakawa, Yi-Hao Peng and Yaxin Hu on earlier drafts of this work.}
\end{acks}
\bibliographystyle{ACM-Reference-Format}
\bibliography{main}

\newpage
\appendix

\section{Participants' Information}
\label{sec:inf}
\begin{figure}[h]
    \centering
        \centering\includegraphics[width=0.7\linewidth]{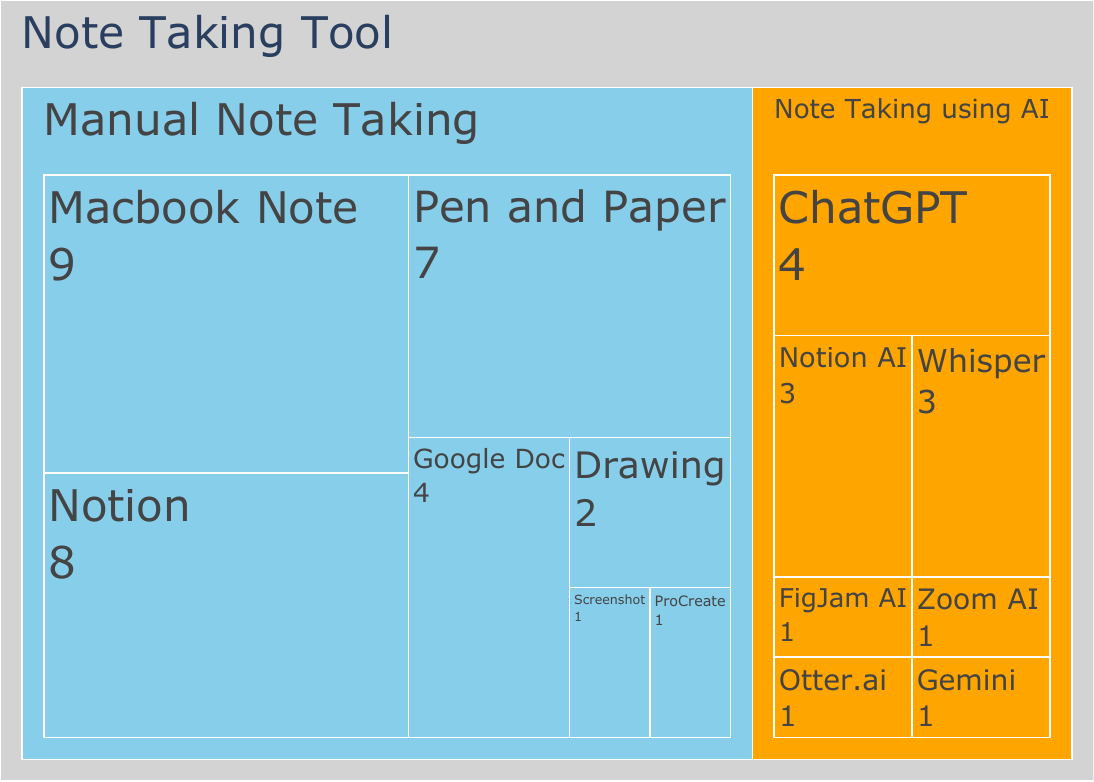}
        \caption{Different tools and techniques participants reported to use for note-taking in their daily lives.}
        \Description{Different tools and techniques participants reported to use for note-taking in their daily lives.}
        \label{fig:tool_use}
\end{figure}

Table \ref{tab:participant-table} shows the demography and occupation of our participants. Table \ref{tab:controlgroup} reports the subgroup division of the participants into four groups. Figure \ref{fig:tool_use} shows the reported tool used by our participants as a tree map.

\begin{table}[h]
\resizebox{\linewidth}{!}{%
\begin{tabular}{llllll}
\hline
\textbf{Group No.} & \textbf{\# of Participants} & \textbf{Tool \#1} & \textbf{Video \#1} & \textbf{Tool \#2} & \textbf{Video \#2} \\ \hline
1 & 3 (P1, P5, P9)  & \tool & \href{https://youtu.be/yOgAbKJGrTA?si=UwlZewxj8v1B54AL}{url1} & Baseline  & \href{https://youtu.be/P7yM0TKvUm4?si=1nSdc4MtTLI3wSKJ}{url2} \\
2 & 3 (P2, P6, P10) & Baseline  & \href{https://youtu.be/yOgAbKJGrTA?si=UwlZewxj8v1B54AL}{url1} & \tool & \href{https://youtu.be/P7yM0TKvUm4?si=1nSdc4MtTLI3wSKJ}{url2} \\
3 & 3 (P3, P7, P11) & \tool & \href{https://youtu.be/P7yM0TKvUm4?si=1nSdc4MtTLI3wSKJ}{url2} & Baseline  & \href{https://youtu.be/yOgAbKJGrTA?si=UwlZewxj8v1B54AL}{url1} \\
4 & 3 (P4, P8, P12) & Baseline  & \href{https://youtu.be/P7yM0TKvUm4?si=1nSdc4MtTLI3wSKJ}{url2} & \tool & \href{https://youtu.be/yOgAbKJGrTA?si=UwlZewxj8v1B54AL}{url1} \\ \hline
\end{tabular}%
}
\caption{Distribution of participants across four subgroups.}
\label{tab:controlgroup}
\end{table}

\begin{table*}[]
\resizebox{\linewidth}{!}{%
\begin{tabular}{lllllllllllll}
 & \textbf{P1} & \textbf{P2} & \textbf{P3} & \textbf{P4} & \textbf{P5} & \textbf{P6} & \textbf{P7} & \textbf{P8} & \textbf{P9} & \textbf{P10} & \textbf{P11} & \textbf{P12} \\ \hline
Baseline & $0.69\pm0.18$ & $0.56\pm0.15$ & $0.55\pm0.17$ & $0.62\pm0.24$ & $0.47\pm0.2$ & $0.60\pm0.15$ & $0.5\pm0.14$ & $0.4\pm0.18$ & $0.57\pm0.16$ & $0.61\pm0.15$ & $0.67\pm0.18$ & $0.53\pm0.18$ \\
NoTeeline & \textbf{$0.86\pm0.05$} & \textbf{$0.84\pm0.04$} & \textbf{$0.84\pm0.04$} & \textbf{$0.83\pm0.07$} & \textbf{$0.86\pm0.07$} & \textbf{$0.83\pm0.06$} & \textbf{$0.73\pm0.19$} & \textbf{$0.86\pm0.05$} & \textbf{$0.83\pm0.11$} & \textbf{$0.84\pm0.06$} & \textbf{$0.85\pm0.04$} & \textbf{$0.86\pm0.04$} 
\\ \hline
\end{tabular}%
}
\caption{Gruen scores of each participant}
\label{tab:gruen_scores}
\end{table*}

\begin{table*}[!ht]
\resizebox{0.8\linewidth}{!}{%
\begin{tabular}{ccccccccccccc}
\multicolumn{1}{l}{} &
  \textbf{P1} &
  \textbf{P2} &
  \textbf{P3} &
  \textbf{P4} &
  \textbf{P5} &
  \textbf{P6} &
  \textbf{P7} &
  \textbf{P8} &
  \textbf{P9} &
  \textbf{P10} &
  \textbf{P11} &
  \textbf{P12} \\ \hline
\textbf{Hallucination Rate (\%)} &
  4.76 &
  5.88 &
  11.11 &
  6.67 &
  6.67 &
  9.09 &
  0.0 &
  0.0 &
  10.71 &
  13.33 &
  10.0 &
  3.33 \\ 
\textbf{Factual Consistency Rate (\%)} &
  95.24 &
  94.12 &
  88.89 &
  93.33 &
  93.33 &
  90.91 &
  100.0 &
  100.0 &
  89.29 &
  86.67 &
  90.0 &
  96.67 \\ \hline
\end{tabular}}
\caption{Correctness of generated note for each participant in terms of Hallucination Rate (\%, lower is better) and Factual Consistency Rate (\%, higher is better)}
\label{tab:hhem}
\end{table*}

\section{Video Selection Criteria}
\subsection{Video Clip Selection for Onboarding Session}
\label{appendix:onboarding_video}
We carefully selected three videos for the onboarding from three different domains: Science, Psychology, and blogs. We extracted 15-20 second snippets from each video and stored the timestamps. The videos are listed below: 1) \href{https://youtu.be/39HTpUG1MwQ?si=HL6HtS-yPt1RtL9-}{How Mitochondria Produce Energy} (Science); 2) \href{https://youtu.be/II5h6uJPvvs?si=8a2DDAohy023RtB6}{How to get motivated even when you don’t feel like it} (Psychology); 3) \href{https://youtu.be/Cf7rYcOZmKg?si=xwICRJjq1VN_O7mS}{Apple Vision Pro VS Meta Quest 3! HONEST Comparison!} (Blog).

\subsection{Video Selection Criteria  for User Study} 
\label{subsec:vid_select}
We picked two videos for the two conditions. The videos met the following criteria: 1) not too long ($<5$ minutes); 2) information heavy where users are exposed to scientific concepts (-- so that they feel the need to take note); 3) interesting to a wide range of audience. Based on these criteria, we selected two TED-Ed videos entitled \textit{`How memories form and how we lose them'} \href{https://youtu.be/yOgAbKJGrTA?si=UwlZewxj8v1B54AL}{(url1)} and \textit{`Is it possible to lose weight fast?'} \href{https://youtu.be/P7yM0TKvUm4?si=1nSdc4MtTLI3wSKJ}{(url2)}. Both videos relate to Human Biology, are \textasciitilde4 minutes long, and have been viewed over 2 million times, reflecting their wide appeal.

If time permitted, we also encouraged participants to try out \tool with a video of their choice to stir personal interest. For this, we explicitly asked participants to freely explore \tool and observed no carry-over biases. Similar ordering of reproduction task and free creation task has been adopted by previous HCI works \cite{Xia2018DataInkDA, xia2016object, datapart}.  

\begin{table*}
\resizebox{\textwidth}{!}{%
\begin{tabular}{ll}
\hline 
\textbf{Expanded Note} &
  \textbf{Summary} \\
  \hline \\
\begin{tabular}[c]{@{}l@{}}\colorbox{table_color_1}{Encoding is done with bidirectional LSTM}, and the \colorbox{table_color_1}{attention mechanism allows} \\ \colorbox{table_color_1}{the decoder to treat encoded representations as a form of memory to reference} \\ and highlight what's important at any time.\end{tabular} &
  \multirow{4}{*}{\begin{tabular}[c]{@{}l@{}}In the lecture, \colorbox{table_color_1}{the use of bidirectional LSTM in encoding is highlighted},\\ \colorbox{table_color_1}{with the attention mechanism enabling the decoder to utilize encoded}\\ \colorbox{table_color_1}{representations as a memory form, focusing on relevant parts at any given} \\ \colorbox{table_color_1}{time}. This approach is foundational in \colorbox{table_color_1}{sequence to sequence models}, particularly\\ in creating \colorbox{table_color_1}{end-to-end differentiable machine translation systems}, where an \\ \colorbox{table_color_1}{encoder and decoder framework} is pivotal. The lecture also touches upon the nature\\ of \colorbox{table_color_1}{recurrent neural networks (RNNs), which can be unrolled in both directions}, thus \\ capturing linear locality and maintaining a computational complexity that is linear with\\ respect to sequence length. \colorbox{table_color_1}{This linear encoding of locality by RNNs} contrasts with the\\ quadratic \colorbox{table_color_1}{computational complexity} challenges faced by transformers, underscoring the\\ evolution and adaptation within NLP technologies.\end{tabular}} \vspace{8pt} \\
\begin{tabular}[c]{@{}l@{}}\colorbox{table_color_1}{Sequence to sequence models} employ \colorbox{table_color_1}{an encoder and decoder framework} \\ to handle tasks like building an end-to-end differentiable machine translation \\ system.\end{tabular} &
   \vspace{8pt} \\
\begin{tabular}[c]{@{}l@{}}Recurrent neural networks \colorbox{table_color_1}{(RNNs) can be unrolled either from left to right or} \\ \colorbox{table_color_1}{from right to left}, encoding linear locality.\end{tabular} &
   \vspace{8pt} \\
\begin{tabular}[c]{@{}l@{}}\colorbox{table_color_1}{Recurrent neural networks encode linear locality with a computational} \\ \colorbox{table_color_1}{complexity} that is linear in terms of the sequence length.\end{tabular} &
   \vspace{8pt} \\
   \hline 
\end{tabular}}
\caption{Expanded notes and summary generated by GPT-4 using these notes. The relevant portions in the expanded notes and summary have been highlighted for readability.}
\label{tab:expand-summary}
\end{table*}

\section{Relevance of expanded note and summary}
Table \ref{tab:expand-summary} shows how the summary is conditioned on user notes and includes the relevant portions from the notes.

\section{Response Time for OpenAI API}
\revision{Table \ref{tab:time-pid} shows the response time for the OpenAI API. On average, OpenAI takes $2$ seconds to generate the expanded notes from micronotes.}

\begin{table}[h]
\resizebox{0.5\linewidth}{!}{%
\begin{tabular}{cc}
\toprule
\textbf{ID} & \begin{tabular}[c]{@{}c@{}}\textbf{API Response Time}\\\textbf{(in seconds)}\end{tabular} \\ \midrule
P1 & 1.83 \\
P2 & 1.47 \\
P3 & 1.53 \\
P4 & 1.43 \\
P5 & 2.25 \\
P6 & 1.75 \\
P7 & 1.71 \\
P8 & 2.67 \\
P9 & 2.61 \\
P10 & 2.09 \\
P11 & 2.37 \\
P12 & 2.38 \\ \midrule
avg. & 2.00 \\ \bottomrule
\end{tabular}%
}
\caption{\revision{API response time (in seconds) for each participant.}}
\label{tab:time-pid}
\end{table}

\section{Prompts used for \tool}

\subsection{Prompt for Note Expansion}
Figure \ref{prompt-expansion} shows the prompt that we use for expanding the micronotes of the user. It specifies some rules explaining what kind of expanded notes are expected. The rules ask to resolve any grammatical mistakes, generate a single sentence as the expanded note point, and avoid some specific phrases to start the expanded sentence. The onboarding transcript, onboarding micronotes, and the onboarding full notes are supplied as examples to the prompt. Moreover, the transcript of the YouTube video and the microntes of the user are also passed to the prompt for expansion.

\subsection{Prompt for Note Organization by Theme}
This prompt can be seen in Figure \ref{prompt-theme}. It explains the structure of the theme that is expected as a response. As an example, some points and their corresponding themes are incorporated into the prompt for better theme generation. Like the note expansion prompt, some rules have also been mentioned here. These rules tell how many points are to be grouped into one theme topic and that all points should be assigned a detailed theme topic.

\subsection{Prompt for Cue Question}
The prompt here in Figure \ref{prompt-cue} asks to generate five multiple-choice questions based on the expanded note that has been provided. It also shows in what format the response is expected. A topic and five multiple-choice questions have been added to the prompt.

\subsection{Prompt for Summary}
The prompt for summary in Figure \ref{prompt-summary} consists of a context and some keypoints. The context here is the preprocessed transcript in Langchain and the keypoints are the expanded note points. It strictly asks to generate a summary in 4 sentences and use the context only for additional information.

\section{Questionnaire for Needfinding Study}
\label{subsec:needfinding_ques}
We asked the participants the following five questions in the Needfinding study:
\begin{itemize}
    \item What is your typical notetaking process when watching online informational videos (e.g., podcasts, lectures, tutorials)?
    \item  What are your main pain points with the current notetaking process?
    \item Do you use any existing AI software to assist with your notetaking? If so, please briefly describe their pros and cons.
    \item Are there any additional notetaking features that you would like to have in the tools you are using now?
    \item If there is an AI-assisted system that could take your manual notes and organize them automatically for you, would you like to use them?
\end{itemize}

\begin{figure*}[ht]
\begin{framed}
{\fontsize{7}{7}
\texttt{I want you to act as a personalized note-taking assistant. Users will give you a keypoint and the YouTube transcript. Your task is to expand the keypoint into a note point, by taking additional context from the transcript. The note should be a full sentence in simple English. Follow these rules: \\ 1. Resolve any typos or grammatical mistakes that arise in the keypoint. \\ 2. The note should not be longer than 1 sentence. \\ 3. Remember that the keypoint can be very abstract and as short as an abbreviation. Use the transcript to get additional information to ensure a good quality note expansion. \\ 4. Just write a single note point, and users will request repeatedly for new points they want to add. \\ 5. Write it in a way a user would write in a notepad. Do not use sentences such as 'This video talks about...', 'The speaker explains..' etc. \\ Make sure that the note aligns with the user's writing style so that they can read it easily. Use the same writing style as shown below. Here are three examples: \\ \textbf{Transcript}: \colorbox{pink}{Onboarding Transcript} \\ \textbf{Keypoint}: \colorbox{yellow}{Onboarding Micronotes} \\ \textbf{Note}: \colorbox{lime}{Onboarding Full Note} \\ The keypoint refers to the high-level keypoint provided by the user and your task is to write a full 'Note' point. Make sure that your expanded note point matches the writing style of 'Note' in the provided examples. Expand the provided keypoint into a one sentence note. \\ \textbf{Transcript}: \colorbox{pink}{Input Transcript} \\ \textbf{Keypoint}: \colorbox{yellow}{Input Micronotes} \\ \textbf{Note}: }}
\end{framed}
\caption{Prompt used for note expansion}
\Description{Prompt used for note expansion}
\label{prompt-expansion}
\end{figure*}

\begin{figure*}[ht]
\begin{framed}
{\fontsize{7}{7}
\texttt{Given a list of points, Your task is to perform topic modeling over them. Arrange the points into topics and provide a name for each topic. Please mark the topic within \textless{}Topic\textgreater\textless{}/Topic\textgreater tags. the points are marked with \textless{}p\textgreater tag as well. Follow these rules: \\ 1. Each topic should have more than one point. \\ 2. Every point should be assigned under one topic. No point should be unassigned. \\ 3. The topic should not be very high-level, make them as detailed as possible. \\ Here is an example: \\ \textbf{Input points}: \\ 1. New York City is home to a vibrant mix of cultures, with communities from nearly every corner of the globe, offering an incredible range of languages, cuisines, and cultural practices. .. \textit{ (cont.)}.. \\ \textbf{Answer}: \\ \textless{}Topic name="Cultural Celebration in New York City"\textgreater \\ \textless{}p\textgreater New York City is home to a vibrant mix of cultures, with communities from nearly every corner of the globe, offering \\ an incredible range of languages, cuisines, and cultural practices.\textless{}/p\textgreater .. \textit{ (cont.)}.. \\ \textless{}/Topic\textgreater \\ \textbf{Input points:} \colorbox{lime}{Expanded Points} \\ \textbf{Answer}:}}
\end{framed}
\caption{Prompt for Note Organization by Theme}
\Description{Prompt for Note Organization by Theme}
\label{prompt-theme}
\end{figure*}

\begin{figure*}[ht]
\begin{framed}
{\fontsize{7}{7}
\texttt{I will give you a context and some keypoints, Your task is to summarize the keypoints in 4 sentences. Focus on the keypoint, only use context if you need extra information: \\ \textbf{Context}: \colorbox{cyan}{Preprocessed Transcript in Langchain} \\ \textbf{Keypoints}: \colorbox{lime}{Expanded Points} \\ remember not to make it too long. Do not mark the sentences with 1, 2, etc.}}
\end{framed}
\caption{Prompt for Summary}
\Description{Prompt for Summary}
\label{prompt-summary}
\end{figure*}

\begin{figure*}[ht]
\begin{framed}
{\fontsize{7}{7}
\texttt{System prompt: \\ Given a topic description, your task is to generate five multichoice questions with answers. Please mark the question within \textless{}Question\textgreater{}\textless{}/Question\textgreater tags, individual choices within \textless{}Choice\textgreater{}\textless{}/Choice/\textgreater tags and answer within \textless{}Answer\textgreater{}\textless{}/Answer\textgreater tags. Make sure not to put the right choice in the same choice option, randomly assign it within A, B, C or D. Here is an example: \\ \textbf{Topic}: Resilience refers to how well you can deal with and bounce back from the difficulties of life. It can mean the difference between handling pressure and losing your cool. Resilient people tend to maintain a more positive outlook and cope with stress .. \textit{ (cont.)}..\\ \textless{}Question\textgreater{}What does resilience refer to?\textless{}/Question\textgreater\\ \textless{}Choice\textgreater{}A. Dealing with difficulties by losing your cool\textless{}/Choice\textgreater\\ \textless{}Choice\textgreater{}B. Bouncing back from the challenges of life\textless{}/Choice\textgreater\\ \textless{}Choice\textgreater{}C. Avoiding stressful situations altogether\textless{}/Choice\textgreater\\ \textless{}Choice\textgreater{}D. Ignoring problems and hoping they go away\textless{}/Choice\textgreater\\ \textless{}Answer\textgreater{}B. Bouncing back from the challenges of life\textless{}/Answer\textgreater\\ User prompt:\\ \textbf{Topic}: \colorbox{lime}{Expanded Note} \\ \textbf{Additional Context to get extra information for the topic}: \colorbox{cyan}{Preprocessed Transcript in Langchain}}}
\end{framed}
\caption{Prompt for Cue Question}
\Description{Prompt for Cue Question}
\label{prompt-cue}
\end{figure*}

\end{document}


\title{NoTeeline: Supporting Real-Time, Personalized Notetaking with LLM-Enhanced Micronotes \\ Supplementary Material}

\author{Faria Huq}
\authornote{Both authors contributed equally to this research.}
\email{fhuq@cs.cmu.edu}
\affiliation{%
  \institution{Carnegie Mellon University}
  \city{Pittsburgh}
  \state{Pennsylvania}
  \country{USA}
}
\author{Abdus Samee}
\authornotemark[1]
\email{1805021@ugrad.cse.buet.ac.bd}
\affiliation{%
  \institution{Bangladesh University of Engr. \& Tech.}
  \city{Dhaka}
  \country{Bangladesh}
}
\author{David Chuan-en Lin}
\email{chuanenl@cs.cmu.edu}
\affiliation{%
  \institution{Carnegie Mellon University}
  \city{Pittsburgh}
  \state{Pennsylvania}
  \country{USA}
}
\author{Xiaodi Alice Tang}
\email{xiaodita@andrew.cmu.edu}
\affiliation{%
  \institution{Carnegie Mellon University}
  \city{Pittsburgh}
  \state{Pennsylvania}
  \country{USA}
}
\author{Jeffrey P. Bigham}
\affiliation{%
  \institution{Carnegie Mellon University}
  \city{Pittsburgh}
  \state{Pennsylvania}
  \country{USA}}
\email{jbigham@cs.cmu.edu}

\renewcommand{\shortauthors}{Trovato and Tobin, et al.}
\newcommand{\todo}[1]{\textcolor{red}{#1}}

\ccsdesc[300]{Human-centered computing~User interface toolkits}
\ccsdesc[300]{Human-centered computing~Natural language interfaces}

\keywords{Large Language Model, Natural Language Interface, Visualization, Personalization, Writing Assistant}

\newcommand{\tool}{NoTeeline\xspace}
\mdfdefinestyle{MyFrame}{%
    roundcorner=3pt,
    linecolor=gray,
    outerlinewidth=0.5pt,
    innertopmargin=5pt,
    innerbottommargin=5pt,
    innerrightmargin=5pt,
    innerleftmargin=5pt,
    backgroundcolor=gray!8!white,
    frametitleaboveskip=\dimexpr-\ht\strutbox\relax,
}

\maketitle

\section{Example Videos participants tried in \tool}
\begin{figure*}[!htp]
\centering
\begin{subfigure}[b]{0.33\linewidth}
  \centering
\includegraphics[width=0.8\linewidth]{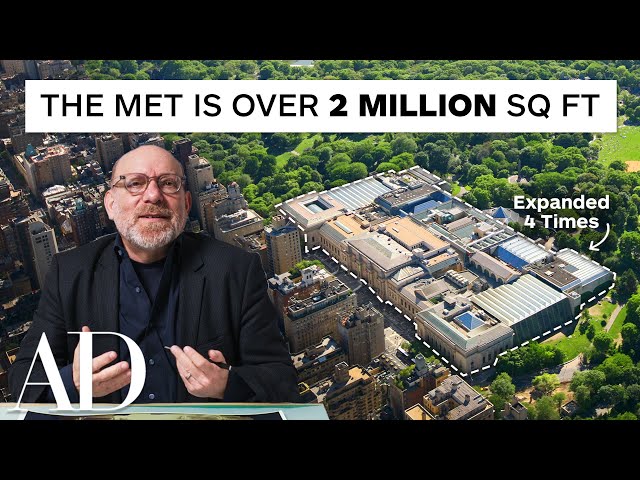}
\caption{Architect Breaks Down The Design of 4 Iconic NYC Museums | Architectural Digest}
\label{fig:p6}
\end{subfigure}
\hfill 
\begin{subfigure}[b]{0.33\linewidth}
  \centering
    \includegraphics[width=0.8\linewidth]{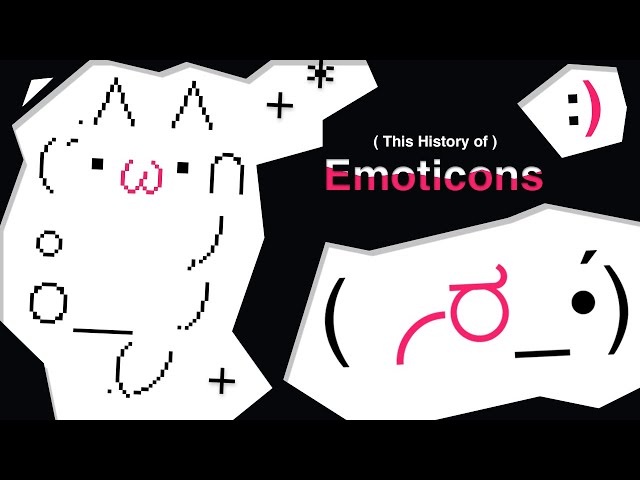}
    \caption{The History of Emoticons}
    \label{fig:p9}
\end{subfigure}
\hfill
\begin{subfigure}[b]{0.33\linewidth}
  \centering
    \includegraphics[width=0.8\linewidth]{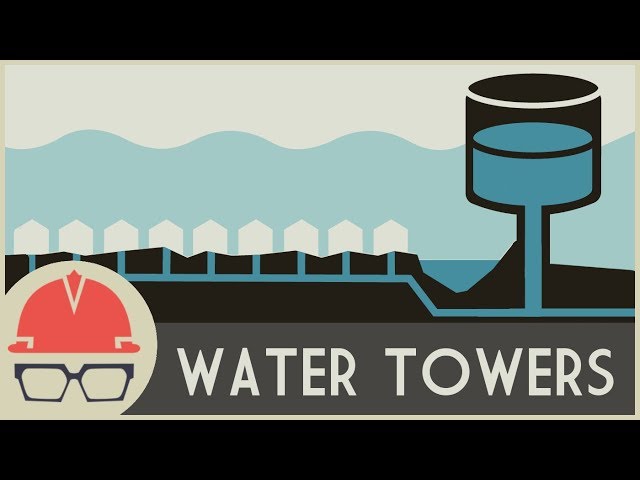}
    \caption{How Water Tower Works}
    \label{fig:p10}
\end{subfigure}
\caption{Examples of videos picked by our participants during the free-form study.} 
\label{fig:threefigures}
\end{figure*}

During the freeform test (\textsection 7.0.3), Participants reported a wide range of use cases where \tool can be useful in their daily lives. We show three examples in Figures \ref{fig:p6}, \ref{fig:p9}, \ref{fig:p10}.

\section{Additional Results}
We show some additional results in this section. Table \ref{tab:supp-summary} shows some example summaries generated from user notes. Table \ref{tab:supp-theme} shows an example of themes generated from the micronotes of user P1 for a video we suggested. Table \ref{tab:supp-comp-notee-base} shows the comparison between the summaries generated from user micornotes by \tool and the web app NoteGPT. Table \ref{tab:supp-quiz} shows the quiz generated from micronotes of user P1 for a video handpicked by that user.

\begin{table*}
\centering
\resizebox{\textwidth}{!}{%
\begin{tabular}{ll}
\multicolumn{1}{c}{\textbf{User}} &
  \multicolumn{1}{c}{\textbf{Summary}} \\
  \hline \\
P1 &
  \begin{tabular}[c]{@{}l@{}}The process of memory formation involves converting experiences into electrical energy that travels through neurons, initially stored in short-term memory \\ before being transferred to long-term memory via the hippocampus. This transfer is facilitated by long-term potentiation, which enhances the efficiency of \\ communication between neurons, aiding in the storage of memories for extended periods. However, factors such as aging, which leads to neuron loss in the \\ hippocampus and weakened synapses, along with mental health issues like chronic stress and depression, can impair memory retrieval and encoding. Engaging \\ in social interactions, staying physically active, and challenging the brain with activities like learning new languages are recommended strategies to slow \\ memory decline and preserve brain health.\end{tabular} \\ \\
P6 &
  \begin{tabular}[c]{@{}l@{}}The Metropolitan Museum of Art's design, deeply influenced by its Victorian Gothic origins, mirrors the fairy-tale essence of its Central Park setting, sharing the \\ same architect, Calvert Vaux, enhancing its historical and aesthetic connection to the surrounding landscape. Over the years, the museum has expanded significantly, \\ now stretching five blocks along Fifth Avenue, while ensuring the original facade remains visible, showcasing a thoughtful blend of growth and preservation. Richard \\ Morris Hunt, known for the base of the Statue of Liberty, contributed to this expansion by adding a classical style to the museum's architecture, enriching its visual \\ and historical narrative. These expansions, including a significant addition by McKim, Mead \& White, have not only allowed for a broader display of its vast collection\\  but also brought the museum closer to the street, making it more accessible and inviting to the public.\end{tabular} \\ \\
P10 &
  \begin{tabular}[c]{@{}l@{}}Water towers play a pivotal role in urban water supply systems by managing demand fluctuations, maintaining pressure, and ensuring a constant and reliable water \\ source for drinking, sanitation, and firefighting. They utilize gravity to store energy and maintain system pressure, reducing the need for continuous pump operation and\\  aiding in public health by preventing contamination during low-pressure situations. To address the challenges of varying elevations within a city, some areas may \\ implement separate pressure zones or use additional pumps in tall buildings. Despite their cost and space requirements, water towers are an essential and dependable \\ component of public water infrastructure, balancing efficiency with the necessity of clean water access for urban populations.\end{tabular} \\ \\
P12 &
  \begin{tabular}[c]{@{}l@{}}BC, a 29-year-old man, was admitted to the emergency room in a disoriented state after consuming 120 probiotic gummies, attempting to counteract the negative \\ effects of antibiotics on his gut health. His overindulgence in gummies, believed to restore gut balance, led to severe health complications including infective \\ endocarditis, likely introduced into his bloodstream through damaged gums. The bacterial infection from the probiotics necessitated surgery to repair his heart \\ valves and a course of targeted antibiotics. This case underscores the risks of excessive consumption of supplements and the importance of moderation, highlighting\\ how beneficial bacteria can turn harmful under certain conditions.\end{tabular} \\
  \hline
\end{tabular}}
\caption{Summary generated from the notes of users}
\label{tab:supp-summary}
\end{table*}


\begin{table*}
\resizebox{\textwidth}{!}{%
\begin{tabular}{ll}
\multicolumn{1}{c}{\textbf{Micronote}}                          & \multicolumn{1}{c}{\textbf{Theme}} \\
\hline \\
1. thinking about vivid memory &
  \multirow{21}{*}{\begin{tabular}[c]{@{}l@{}} \textbf{Theme} name="Memory Formation and Retention Processes" \\\phantom{1pt} {}experience is converted to electricity \\\phantom{1pt} {}goes to short term memory \\\phantom{1pt} {}then goes to long term, hippocampus \\\phantom{1pt} {}neurons communicate with synapses \\\phantom{1pt} {}they get better at communicating \\\phantom{1pt} {}long-term potentiation \\  \textbf{Theme} name="Memory Strength and Fading" \\\phantom{1pt} {}thinking about vivid memory \\\phantom{1pt} {}then thinking about lunch from a while ago \\\phantom{1pt} {}why do memories fade? \\  \textbf{Theme} name="Effects of Aging and Health on Memory" \\\phantom{1pt} {}age makes synapses worse \\\phantom{1pt} {}hippocampus shrink \\\phantom{1pt} {}20\% loss by 80 \\\phantom{1pt} {}depression also makes memory formation hard \\\phantom{1pt} {}low serotonin at fault \\\phantom{1pt} {}isolation, thinking about sad memories linked \\  \textbf{Theme} name="Strategies for Memory Preservation" \\\phantom{1pt} {}memories encoded when it's menaningful and we're paying attention \\\phantom{1pt} {}chronic stress makes this hard \\\phantom{1pt} {}social interaction works out the brain \\\phantom{1pt} {}how to keep memories? \\\phantom{1pt} {}eat well and work out \\\phantom{1pt} {}give your brain a workout helps too {}\end{tabular}} \\  
2. then thinking about lunch from a while ago                        &                                    \\
3. why do memories fade?                                             &                                    \\
4. experience is converted to electricity                            &                                    \\
5. goes to short term memory                                         &                                    \\
6. then goes to long term, hippocampus                               &                                    \\
7. neurons communicate with synapses                                 &                                    \\
8. they get better at communicating                                  &                                    \\
9. long-term potentiation                                            &                                    \\
10. age makes synapses worse                                          &                                    \\
11. hippocampus shrinks                                               &                                    \\
12. 20\% loss by 80                                                   &                                    \\
13. memories encoded when it's menaningful and we're paying attention &                                    \\
14. chronic stress makes this hard                                    &                                    \\
15. depression also makes memory formation hard                       &                                    \\
16. low serotonin at fault                                            &                                    \\
17. isolation, thinking about sad memories linked                     &                                    \\
18. social interaction works out the brain                            &                                    \\
19. how to keep memories?                                             &                                    \\
20. eat well and work out                                             &                                    \\
21. give your brain a workout helps too                               &    
\vspace{2cm}
\\ \hline
\end{tabular}}
\caption{Themes with note points written by a user with id P1}
\label{tab:supp-theme}
\end{table*}

\begin{table*}
\resizebox{0.8\textwidth}{!}{%
\begin{tabular}{ll}
\hline
\textbf{User Micronote} &
  \begin{tabular}[c]{@{}l@{}}1. Word vectors -  example with color vector\\ 2. RGB points for each - which color name similar color.\\ 3. Find the color name's average\\ 4. used python\\ 5. \colorbox{table_color_5}{subtracting / adding vector}\\ 6. \colorbox{table_color_5}{find the average / adding vectors together}\\ 7. closest color name in RGB - ucline distance in vector space : find the neighbor color\\ 8. ex. five word -  what is the closest word\\ 9. red -  all similar reds\\ 10. ex 2. with RGB \\ 11. \#\#\#\# operate on language and numbers\\ 12. ex. purple -red -  it finds blue in vector space\\ 13. ex2. \colorbox{table_color_5}{add blue to green it finds cyan}\\ 14. ex3. \colorbox{table_color_5}{average black and white and generate grey}\end{tabular} \\ \vspace{0.5pt} \\
\textbf{NoTeeline Summary} &
  \begin{tabular}[c]{@{}l@{}}Word vectors function similarly to color vectors in representing concepts within a space, where \\ RGB values are used to denote specific colors, akin to naming colors. \colorbox{table_color_5}{By calculating the mean} \\ value of different color names or vectors, one can find the average color, which involves \colorbox{table_color_5}{adding} \\ \colorbox{table_color_5}{vectors together and then finding the mean value to determine a resultant color}. Manipulating the \\ direction and magnitude of these vectors through \colorbox{table_color_5}{addition or subtraction} allows for the creation \\ of new vectors, representing the blending or differentiation of colors. For example, mixing red \\ and blue vectors finds blue in vector space, while \colorbox{table_color_5}{combining blue and green yields cyan}, \\ and \colorbox{table_color_5}{averaging black and white creates grey}.\end{tabular} \\ \vspace{0.5pt} \\
\textbf{NoteGPT Summary} &
  \begin{tabular}[c]{@{}l@{}}Allison Parrish discusses experimental creative writing with vectorized word, explaining \\ vector representations of semantics and phonetics, and demonstrates their use in poetry generation.\end{tabular} \\ \hline
\end{tabular}}
\caption{Comparison of baseline (NoteGPT) and NoTeeline summary. Compared to the baseline, the NoTeeline summary is more aligned with the notes.}
\label{tab:supp-comp-notee-base}
\end{table*}

\begin{table*}
\resizebox{\textwidth}{!}{%
\begin{tabular}{ll}
\multicolumn{1}{c}{\textbf{Note Points}} &
  \multicolumn{1}{c}{\textbf{Quiz}} \\
\hline \\
\begin{tabular}[c]{@{}l@{}}1. Chieko introduces her AI suitcase project\\ 2. Cites biggest fear with losing independence\\ 3. I suitcase to help navigate the world.\\ 4. Uses IBM Watson as the underlying tech\\ 5. Cites difficulties for BVI people\\ 6. Information accessibility.\\ 7. machine was hard to use and nosiy\\ 8. Mobility\\ 9. hard to get to classes or where you want when you want\\ 10. Joined IBM at 1985\\ 11. Introduces IBM's history with people with disabilities\end{tabular} &
  \begin{tabular}[c]{@{}l@{}} \textbf{Question.} {}What is the primary purpose of the AI Suitcase introduced by Chieko Asakawa? \\    {}A. To serve as a fashionable accessory for the visually impaired   \\    {}B. To assist visually impaired individuals in navigating complex environments   \\    {}C. To store and manage personal items with AI technology   \\    {}D. To act as a portable computer for business purposes   \\  \textbf{Answer} {}B. To assist visually impaired individuals in navigating complex environments\\ \\  \textbf{Question.} {}Which technology does the AI Suitcase primarily utilize to assist users? \\    {}A. IBM Watson   \\    {}B. Microsoft Azure   \\    {}C. Google Cloud   \\    {}D. Amazon Web Services   \\  \textbf{Answer} {}A. IBM Watson\\ \\  \textbf{Question.} {}What was Chieko Asakawa's biggest fear that led to the creation of the \\ AI Suitcase project? \\    {}A. The fear of technology becoming too advanced   \\    {}B. The fear of losing personal items while traveling   \\    {}C. The fear of losing independence   \\    {}D. The fear of not being able to keep up with technological advancements   \\  \textbf{Answer} {}C. The fear of losing independence \\ \end{tabular} \\
  \hline
\end{tabular}}
\caption{Quiz generated from the note points of participant, P1}
\label{tab:supp-quiz}
\end{table*}